# Feasibility of observing the α decay chains from isotopes of SHN with Z = 128, Z = 126, Z = 124 and Z = 122

K. P. Santhosh*, B. Priyanka and C. Nithya

*School of Pure and Applied Physics, Kannur University, Swami Anandatheertha Campus, Payyanur 670327, Kerala, India*

**Abstract**

Taking the Coulomb and proximity potential model for deformed nuclei (CPPMDN) as the interacting potential for the post-scission region, the alpha decay properties of 34 isotopes of the superheavy nuclei with Z = 128 within the range 306 ≤ A ≤ 339 have been studied, considering both the parent and daughter nuclei to be deformed. The manuscript also deals with the decay properties of the isotopes of Z = 126 (within 288 ≤ A ≤ 339), Z = 124 (within 284 ≤ A ≤ 339) and Z = 122 (within 280 ≤ A ≤ 339). The alpha decay half lives thus evaluated has been compared with the values evaluated using other theoretical models and it was seen that, our theoretical alpha decay half lives matches well with these values. Through the present study, we have underlined and have established the fact that, among the 192 isotopes considered in the present study, only those isotopes $^{321-324,328-335}$128, $^{318-320,323-327}$126, $^{305-308,315-322}$124 and $^{298-307,311-314}$122 can be synthesized and detected through alpha decay in laboratory. As the alpha decay half lives of these superheavy isotopes lie within the experimental limits, we hope these predictions, on the decay modes of these unknown nuclei, to pave the way for the future experiments. The proton separation energy calculations on $^{306-339}$128, $^{288-336}$126, $^{284-339}$124 and $^{280-339}$122 superheavy nuclei have also been done and the study revealed the probable proton emitters among these nuclei.

*email: drkpsanthosh@gmail.com

## 1. Introduction

The experimental advancements and the recent experimental results [1-11] have evinced the possibilities in the production and investigation of nuclei in the so-called region, the "island of stability/magic island". The notion on the existence of superheavy nuclei (SHN) was raised in the middle of the 1960s by Myers and Swiatecki [12] and later on the search for new isotopes in the superheavy (SH) region have grown as a piquant topic in nuclear physics. The authors concluded that, to expect a superheavy nucleus, both the proton and neutron (spherical) shells of the corresponding nucleus should be closed and thereby the nucleus and its neighbours could have half lives long enough to be observed. Those nuclei devoid of these shell effects would immediately decay owing to the large Coulomb repulsion between protons (large Z), as indicated by macroscopic models. The effects of shell structure are important for all nuclei. However, the role of shell effects for the heaviest ones is more crucial, as these nuclei, in particular almost all transactinide nuclei, would not exist without these effects [13].

The production process of SHN has been performed mainly through the fusion of heavy nuclei just above the barrier [10]. The cold fusion reactions [10], with the double magic nucleus $^{208}$Pb or nearly magic nucleus $^{209}$Bi targets, and the hot fusion reactions [14], with the double magic nucleus $^{48}$Ca bombarding an actinide nucleus in a fusion-evaporation reaction mechanism, are the two processes by which the SHN can be synthesised. The experimentalists at the Joint Institute for Nuclear Research-

Flerov Laboratory of Nuclear Reactions (JINR-FLNR), Dubna, in collaboration with the Lawrence Livermore National Laboratory researchers have been successful in the synthesis of SHN with Z = 113-118 [2-5] and an attempt on the production of Z = 120 [1] through the hot fusion reactions. The cold fusion reactions which paved a way for the synthesis of SHN with Z = 107-112 at GSI, Darmstadt [7, 9, 10, 15, 16] have also been performed at RIKEN, Japan, through which the experimentalists could successfully synthesise the isotopes of SHN with Z = 113 [6] and could also reconfirm [8, 17] the existence of the superheavy elements Z = 110, 111, and 112 reported earlier by the GSI group. Thus, although the history on the progress in the production of SHN, as a distinct realm of nuclear physics, is rather short, the results are quite rich and remarkable. Theoretically, at present, those nuclei which exist due to their shell structure [9, 18, 19] can be regarded to as SHN and as the description of shell structure and its effects on half lives depends on the approach used, the definition is not sharp. However, all realistic descriptions indicate that SHN corresponds roughly to those nuclei with $Z \geq 104$, nuclei of transactinide elements. Thus, as per the above definition, synthesis of about 85 SHN with Z = 104-118, i.e., of 15 SHE has already been reported.

The disclosures of new elements with higher Z and the estimations on their decay properties have always provided deep insights into the understanding of the behaviour of nuclear matter under extreme conditions of large Z and have also helped in predicting an island of stability around N = 184 and Z around 114 or 120-126. A more detailed information on the nuclear structure can be obtained through the synthesis of odd-Z SHN, because of their longer alpha decay chains as a result of strong fission hindrance caused by the unpaired nucleons. Small cross sections for synthesis of SHN (generally below nanobarns) and, simultaneously, short half-lives (generally below seconds) can be pointed out as the major hindrances in synthesising new SHN.

Many of the SHN decay through alpha emission followed by spontaneous fission. Using parabolic potential approximation, Xu et al., [20] has proposed a semi empirical formula for calculating the spontaneous fission half lives of SHN. The agreement between theoretical and experimental results were satisfactory while using the semi empirical formula of Xu et al. Spontaneous fission half life calculations of 160 heavy and superheavy nuclei has been done by Warda et al., [21] within the Hartree-Fock-Bogoliubov (HFB) approach with the finite range and density dependent Gogny force with the DIS parameter set. The study shows reasonably good agreement with the experimental data. Staszczak et al., [22] calculated the spontaneous fission modes and life times of superheavy elements in the nuclear density functional theory. The study gives a systematic self-consistent approach to spontaneous fission in SHN. Most of the SHN synthesized until now are observed via the α cascade decay [1-11, 14-17, 23]. In quantum theory, the process of α decay is considered to be a potential barrier-penetration problem and thus the WKB method can be used to describe the alpha decay half lives. The height of the potential barrier and the position of the barrier in the potential are the decisive factors in describing the α decay half lives. Presently, several theoretical approaches like the cluster model [24], fission model [25], the density-dependent M3Y (DDM3Y) effective interaction [26], the generalized liquid drop model (GLDM) [27], which comes under the broad area of macro-micro method, and self consistent theories like the relativistic mean field theory [28] and the Skyrme-Hartree-Fock mean field model [29] are being used to describe the alpha decay from heavy and SHN. Several properties of nuclei like the deformations, masses, α decay energies, rotational properties, fission barriers, half lives, single-particle levels, exotic geometry of SHN and role of shell effects could also be explained using the theoretical approaches mentioned above. Systematic calculations on the α decay energies ($Q_\alpha$) and α decay half-lives of superheavy nuclei (SHN) with $Z \geq 100$ are performed by using 20 mass models and 18 empirical formulas respectively by Wang et al [30]. From the calculated values of the average deviation and standard deviation for 121 SHN, the authors found that WS4 mass model [31] is the most accurate one to reproduce the experimental $Q_\alpha$ values of the SHN. Among 18 formulae used to calculate the α decay half lives, this study shows that SemFIS2 formula [32] is the best one to predict α decay half-lives. In addition UNIV2 formula [32] with fewer parameters and the VSS [33, 34], SP [35, 36] and NRDX [37] formulae with fewer parameters work well in predicting the α decay half-lives of SHN. In spite of these approaches, the Coulomb and proximity potential model (CPPM) [38] proposed by Santhosh et al., in 2000 and the Coulomb and proximity potential model for deformed nuclei (CPPMDN) [39], the modified version of CPPM, proposed by Santhosh et al., in 2011, both of which comes under the broad class of fission models, are very successful in describing the process of fission [40, 41], the α particle emission [42-47] and the heavy particle decay [48]. A very recent study has also been done on the competition between the alpha decay and spontaneous fission in odd-even and odd-odd nuclei in the range $99 \leq Z \leq 129$ [49].

Recently, the emission of clusters heavier than α particle from the superheavy nuclei has received much attention. The calculations on the heavy particle radioactivity (HPR) for SHN with Z = 104-124 by Poenaru et al., [50] revealed the possibility of finding regions in which HPR is stronger than alpha decay. A trend towards shorter half-lives and larger branching ratio of HPR relative to α decay can be seen from these studies.

Several theoretical studies have been performed to identify the various properties of SHN and most of these studies indicated that the reasonable candidates for magic numbers (closed shells), next to the experimentally known Z = 82 and N = 126, could be Z = 114 and N = 184 [51, 52], corresponding to the nucleus $^{298}$114. As a consequence of these predictions, the decay half lives of the nuclei around $^{298}$114 were performed [53, 54] and as most of the results were quite optimistic, a search for the existence of new superheavy elements was ignited, which resulted in furthermore exhilarating works on the synthesis of SHN at various laboratories [55, 56].

Even though certain theoretical studies could be quoted to have been performed on the cluster and alpha decays of heavy and superheavy nuclei leading to Z = 126 or N = 126, the behaviour of the isotopes of SHN with $Z \geq 126$ against alpha decay [57] remains unknown. Being the next predicted magic number, we were interested in studying the alpha decay properties of the isotopes of those nuclei around Z = 126 and the objective of the present manuscript is to forecast the mode of decay of 34 isotopes of the yet-to-be synthesised SHN with Z = 128 and its decay products. Through the manuscript, we have also explored the decay properties and the mode of decay of the isotopes of 47 isotopes of Z = 126, 53 isotopes of Z = 124 and 58 isotopes of Z = 122. The present work has been done as an extension to our earlier works on the alpha decay properties and mode of decay of SHN with Z = 115 [42], Z = 117 [43, 44], Z = 118 [45], Z = 119 [46] and Z = 120 [47], whereby we have proved the reliability of CPPMDN in predicting the mode of decay of SHN. The idea of the existence of deformed SHN [34] stimulated us to study the effect of deformation on the alpha decay half lives and thus in addition to the consideration of spherical parent and daughter isotopes, we have also included the deformation values (of both parent and daughter isotopes) and have reviewed the alpha decay properties.

The minutiae of the model used for the present calculations have been presented in section 2, discussion on the alpha decay and spontaneous fission of the nuclei under study and the results obtained have been given in section 3 and in section 4 a conclusion on the entire work has been given.

## 2. The Coulomb and proximity potential model for deformed nuclei (CPPMDN)

In CPPMDN, the emitted cluster (here the alpha particle) is assumed to be spherical but the parent and daughter nuclei may have axially symmetric deformation and the potential energy barrier will depend on the polar angle $\theta$ between the axis of symmetry of the parent or daughter and the direction of alpha particle. The interacting potential between two nuclei in CPPMDN is taken as the sum of deformed Coulomb potential, deformed two term proximity potential and centrifugal potential, for both the touching configuration and for the separated fragments. For the pre-scission (overlap) region, simple power law interpolation as done by Shi and Swiatecki [58] has been used.

The interacting potential barrier for two spherical nuclei is given by

$$V = \frac{Z_1 Z_2 e^2}{r} + V_P(z) + \frac{\hbar^2 \ell(\ell+1)}{2\mu r^2}, \quad \text{for } z > 0 \tag{1}$$

where $Z_1$ and $Z_2$ are the atomic numbers of the daughter and emitted cluster, '$r$' is the distance between fragment centres, '$z$' is the distance between the near surfaces of the fragments, $\ell$ represents the angular momentum, $\mu$ the reduced mass and $V_P$ is the proximity potential. Shi and Swiatecki [58] were the first to use the proximity potential in an empirical manner and later on, several theoretical groups [59-61] have used the proximity potential, quite extensively for various theoretical studies. The contribution of both the internal and the external part of the barrier has been considered, in the present model, for the penetrability calculation and the assault frequency, $v$ is calculated for each parent-cluster combination which is associated with the vibration energy. However, for even $A$ parents and for odd $A$ parents, Shi and Swiatecki [62] get $v$ empirically, unrealistic values as $10^{22}$ and $10^{20}$, respectively.

The proximity potential, $V_P$, of Blocki et al., [63, 64] is given as,

$$V_p(z) = 4\pi\gamma b \left[\frac{C_1 C_2}{(C_1 + C_2)}\right] \Phi\left(\frac{z}{b}\right) \tag{2}$$

with the nuclear surface tension coefficient,

$$\gamma = 0.9517[1 - 1.7826(N-Z)^2/A^2] \text{ MeV/fm}^2 \tag{3}$$

Here $N$, $Z$ and $A$ represent the neutron, proton and mass number of the parent and $\Phi$ represents the universal proximity potential [64] given as

$$\Phi(\varepsilon) = -4.41 e^{-\varepsilon/0.7176}, \text{ for } \varepsilon > 1.9475 \tag{4}$$

$$\Phi(\varepsilon) = -1.7817 + 0.9270\varepsilon + 0.0169\varepsilon^2 - 0.05148\varepsilon^3, \text{ for } 0 \leq \varepsilon \leq 1.9475 \tag{5}$$

with $\varepsilon = z/b$, where the width (diffuseness) of the nuclear surface $b \approx 1$ fermi and the Süsmann central radii $C_i$ of the fragments are related to the sharp radii $R_i$ as

$$C_i = R_i - \left(\frac{b^2}{R_i}\right) \tag{6}$$

For $R_i$, we use semi-empirical formula in terms of mass number $A_i$ as [63]

$$R_i = 1.28 A_i^{1/3} - 0.76 + 0.8 A_i^{-1/3} \tag{7}$$

For the internal part (overlap region) of the barrier, the potential is given as,

$$V = a_0 (L - L_0)^n, \text{ for } z < 0 \tag{8}$$

where $L = z + 2C_1 + 2C_2$ and $L_0 = 2C$, the diameter of the parent nuclei. The constants $a_0$ and $n$ are determined by the smooth matching of the two potentials at the touching point.

Using the one dimensional Wentzel-Kramers-Brillouin approximation, the barrier penetrability $P$ is given as

$$P = \exp\left\{-\frac{2}{\hbar}\int_a^b \sqrt{2\mu(V-Q)}\, dz\right\} \tag{9}$$

The turning points "$a$" and "$b$" are determined from the equation, $V(a) = V(b) = Q$, the energy released. In eqn. (9), the mass parameter is replaced by $\mu = m A_1 A_2 / A$, where $m$ is the nucleon mass and $A_1$, $A_2$ are the mass numbers of daughter and emitted cluster respectively. The above integral can be evaluated numerically or analytically, and the half life time is given by

$$T_{1/2} = \left(\frac{\ln 2}{\lambda}\right) = \left(\frac{\ln 2}{\nu P}\right) \tag{10}$$

where, $\nu = \left(\frac{\omega}{2\pi}\right) = \left(\frac{2E_\nu}{h}\right)$ represent the number of assaults on the barrier per second and $\lambda$ the decay constant. $E_\nu$, the empirical vibration energy is given as [65]

$$E_\nu = Q\left\{0.056 + 0.039 \exp\left[\frac{(4 - A_2)}{2.5}\right]\right\}, \quad \text{for } A_2 \geq 4 \tag{11}$$

Classically, the $\alpha$ particle is assumed to move back and forth in the nucleus and the usual way of determining the assault frequency is through the expression given by $\nu$ = velocity/(2R), where $R$ is the radius of the parent nuclei. As the alpha particle has wave properties, a quantum mechanical treatment is more accurate. Thus, assuming that the alpha particle vibrates in a harmonic oscillator potential with a frequency $\omega$, which depends on the vibration energy $E_\nu$, we can identify this frequency as the assault frequency $\nu$ given in eqns. (10) and (11).

The Coulomb interaction between the two deformed and oriented nuclei with higher multipole deformation included [66, 67] is taken from Ref. [68] and is given as,

$$V_C = \frac{Z_1 Z_2 e^2}{r} + 3 Z_1 Z_2 e^2 \sum_{\lambda, i=1,2} \frac{1}{2\lambda+1} \frac{R_{0i}^\lambda}{r^{\lambda+1}} Y_\lambda^{(0)}(\alpha_i) \left[\beta_{\lambda i} + \frac{4}{7}\beta_{\lambda i}^2 Y_\lambda^{(0)}(\alpha_i)\delta_{\lambda,2}\right] \tag{12}$$

with

$$R_i(\alpha_i) = R_{0i}\left[1 + \sum_\lambda \beta_{\lambda i} Y_\lambda^{(0)}(\alpha_i)\right] \tag{13}$$

where $R_{0i} = 1.28A_i^{1/3} - 0.76 + 0.8A_i^{-1/3}$. Here $\alpha_i$ is the angle between the radius vector and symmetry axis of the $i^{th}$ nuclei (see Fig.1 of Ref [66]) and it is to be noted that the quadrupole interaction term proportional to $\beta_{21}\beta_{22}$, is neglected because of its short-range character.

The proximity potential and the double folding potential can be considered as the two variants of the nuclear interaction [69, 70]. In the description of interaction between two fragments, the latter is found to be more effective. The proximity potential of Blocki *et al.*, [63, 64], which describes the interaction between two pure spherically symmetric fragments, has one term based on the first approximation of the folding procedure and the two-term proximity potential of Baltz *et al.*, (equation (11) of [71]) includes the second component as the second approximation of the more accurate folding procedure. The authors have shown that the two-term proximity potential is in excellent agreement with the folding model for heavy ion reaction, not only in shape but also in absolute magnitude (see figure 3 of [71]). The two-term proximity potential for interaction between a deformed and spherical nucleus is given by Baltz *et al.*, [71] as

$$V_{P2}(R,\theta) = 2\pi \left[\frac{R_1(\alpha)R_C}{R_1(\alpha)+R_C+S}\right]^{1/2} \left[\frac{R_2(\alpha)R_C}{R_2(\alpha)+R_C+S}\right]^{1/2}$$

$$\times \left[\left[\varepsilon_0(S) + \frac{R_1(\alpha)+R_C}{2R_1(\alpha)R_C}\varepsilon_1(S)\right]\left[\varepsilon_0(S) + \frac{R_2(\alpha)+R_C}{2R_2(\alpha)R_C}\varepsilon_1(S)\right]\right]^{1/2} \tag{14}$$

where $R_1(\alpha)$ and $R_2(\alpha)$ are the principal radii of curvature of the daughter nuclei at the point where polar angle is $\alpha$, $R_C$ is the radius of the spherical cluster, $S$ is the distance between the surfaces along the straight line connecting the fragments and $\varepsilon_0(S)$ and $\varepsilon_1(S)$ are the one dimensional slab-on-slab function.

### 3. Results and discussion

A detailed theoretical study, on the alpha decay half lives of the isotopes of SHN with Z = 128 within the range 306 ≤ A ≤ 339, Z = 126 within the range 288 ≤ A ≤ 339, Z = 124 within the range 284 ≤ A ≤ 339 and Z = 122 within the range 280 ≤ A ≤ 339, and the successive decay products of these nuclei, has been performed using the Coulomb and proximity potential model for deformed nuclei (CPPMDN), so that the mode of decay of these respective isotopes could be identified. As referred earlier, even though several experimental studies has been performed by Oganessian et al., aiming at the synthesis of superheavy isotopes with long alpha half lives, any evidence on the synthesis of elements beyond Z > 118 is unknown, except for an attempt on synthesizing Z = 120 [1]. The present study aims at studying the behavior of the isotopes of the SHN, with and around Z = 126, a predicted magic number, against alpha decay, spontaneous fission and proton decay and there by confer a theoretical prediction on the mode of decay of these isotopes.

The alpha transitions between the ground state energy levels of the parent nuclei and the ground state energy levels of the daughter nuclei involves an energy release referred to as the $Q$ value of the reaction and is given as,

$$Q_{gs \to gs} = \Delta M_p - (\Delta M_\alpha + \Delta M_d) + k(Z_p^\varepsilon - Z_d^\varepsilon) \tag{15}$$

The terms $\Delta M_p, \Delta M_d$ and $\Delta M_\alpha$ in the above equation represents the mass excess of the parent, daughter and alpha particle respectively. In the heavy mass region, the alpha decay half lives are extremely sensitive to $Q$ value of the reaction, and an uncertainty of 1MeV in $Q$ value corresponds to an uncertainty of α decay half lives ranging from $10^3$ to $10^5$ times. Hence the $Q$ value calculations for the half lives of α decay has to be considered substantive. In the present manuscript, four different mass tables have been used mainly for calculating the $Q$ values for the alpha decay, namely the recent experimental mass table of Wang *et al.*, [72] and the theoretical mass tables of Koura-Tachibana-Uno-Yamada (KTUY) [73], Kowal et al., [74] and Moller et al., [75]. The mass excess values taken from the Ref. [72] has been used for most of the nuclei under study, and for those nuclei whose experimental mass excess were unavailable, the corresponding values were obtained from the Ref. [73-75]. Here, we would also like to mention that, as the mass excess values of the even-odd isotopes $^{289,291,293,295,297}$126, $^{285,287,289}$124 and $^{281,283}$122 were unavailable from any of these mass tables, these isotopes have not been considered in the present study. The effect of the atomic electrons on the energy of the alpha particle has not been included in the mass excess given in Ref. [72-75]. Hence, to incorporate the electron screening effect [76] and for an accurate calculation of $Q$ value, the term $k(Z_p^\varepsilon - Z_d^\varepsilon)$ has been included in eqn. (15). Here, $k = 8.7$eV and $\varepsilon = 2.517$ for nuclei with Z ≥ 60 and $k = 13.6$eV and $\varepsilon = 2.408$ for nuclei with Z < 60. The ground state deformations of both the parent nuclei and the daughter nuclei have also been incorporated in the present work for the calculation of alpha half lives through the quadrupole ($\beta_2$) and hexadecapole ($\beta_4$) deformation values and as the experimental deformation values were unavailable for the considered nuclei, the corresponding theoretical values were taken from Ref. [77].

**3.1 Alpha half lives of superheavy nuclei**

Three different theoretical models, namely the analytical formulae of Royer [78], the Viola-Seaborg semi-empirical relationship (VSS) [33] and the Universal (UNIV) curve of Poenaru et al., [79, 80] have also been used for evaluating the alpha decay half lives of all the isotopes under study. This has helped in a theoretical comparison of the compatibility of our predicted alpha decay half lives with these theoretical models, and the formalisms are discussed below.

**3.1.1 The analytical formulae of Royer**

Several expressions were advanced for evaluating the decay half lives since the earliest formalism of Geiger and Nutall [81]. Recently, Royer determined the potential energy governing α emission within the liquid drop model including the proximity effects between the α particle and the daughter nucleus and the α decay half lives were deduced from the WKB barrier penetration probability as for a spontaneous asymmetric fission. The theoretical predictions for the heavy and SHN could be well presented using the simple analytical formulae developed by Royer [78]. On applying a fitting procedure to a set of 373 alpha emitters, the following formula was developed with an RMS deviation of 0.42, given as,

$$\log_{10}[T_{1/2}(s)] = -26.06 - 1.114 A^{1/6}\sqrt{Z} + \frac{1.5837 Z}{\sqrt{Q_\alpha}} \qquad (16)$$

Here *A* and *Z* represents respectively the mass number and charge number of the parent nuclei and $Q_\alpha$ represents the energy released during the reaction. Assuming the same dependence on *A*, *Z* and experimental $Q_\alpha$, equation (16) was reformulated for a subset of 131 even-even nuclei and a relation was obtained with a RMS deviation of only 0.285, given as,

$$\log_{10}[T_{1/2}(s)] = -25.31 - 1.1629 A^{1/6}\sqrt{Z} + \frac{1.5864 Z}{\sqrt{Q_\alpha}} \qquad (17)$$

For a subset of 106 even-odd nuclei, the relation given by equation (16) was further modified with an RMS deviation of 0.39, and is given as,

$$\log_{10}[T_{1/2}(s)] = -26.65 - 1.0859 A^{1/6}\sqrt{Z} + \frac{1.5848 Z}{\sqrt{Q_\alpha}} \qquad (18)$$

The eqn. (16) was also reformulated for a subset of 86 odd-even nuclei and 50 odd-odd nuclei. But, in the present study, as we have considered only the even-even and even-odd nuclei, the eqns. (17) and (18) were only used.

### 3.1.2 The Viola-Seaborg semi-empirical relationship (VSS)

Through a systematic analysis of the experimental half lives for alpha decay of heavy elements with A > 140, Viola and Seaborg derived a semi empirical relation [33] which could be used for the prediction of half lives of undiscovered nuclides. Based on a square well nuclear model, the α decay half lives can be expressed as,

$$\log_{10}(T_{1/2}) = A_Z Q_{eff}^{-1/2} + B_Z + \log F \qquad (19)$$

where the $T_{1/2}$ is expressed in seconds, $Q_{eff}$ is the effective α decay energy inside the nucleus in MeV, the constants $A_Z$ and $B_Z$ are the Z dependent coefficients to be determined from fitting the experimental data and log *F* is the hindrance factor for nuclei with unpaired nucleons [82].

Later, by readjusting the original parameters to take into account the new data for even-even nuclei, Sobiczewski, Patyk and Cwiok re-formulated [34] the Viola-Seaborg semi-empirical relationship (VSS) and is given as,

$$\log_{10}(T_{1/2}) = (aZ+b)Q_\alpha^{-1/2} + cZ + d + h_{\log} \qquad (20)$$

where *Z* is the atomic number of the parent nucleus, $T_{1/2}$ is in seconds, $Q_\alpha$ is in MeV and *a*, *b*, *c*, *d* are adjustable parameters. The term $h_{\log}$ which replaced the term log *F* of eqn. (19) represents the hindrances associated with odd proton and odd neutron numbers, as given by Viola-Seaborg [33]. In the present manuscript, instead of using the original set of constants given by Viola and Seaborg [33], the values determined by Sobiczewski et al., [34] have been used. The constants thus used here are *a* = 1.66175, *b* = -8.5166, *c* = -0.20228, *d* = -33.9069 and

$$h_{\log} = \begin{cases} 0, & \text{for } Z = even \quad N = even \\ 0.772, & \text{for } Z = odd \quad N = even \\ 1.066, & \text{for } Z = even \quad N = odd \\ 1.114, & \text{for } Z = odd \quad N = odd \end{cases} \quad (21)$$

### 3.1.3 The Universal curve (UNIV) of Poenaru et al.,

Poenaru et al., while analysing the cluster radioactivity, showed that the preformed cluster models (PCM) are equivalent with fission models used to describe the cluster radioactivities and alpha decay in a unified manner. The authors interpreted cluster pre-formation probability as the penetrability of the pre-scission part of the barrier for the first time, and developed a linearized universal curve (UNIV) [83, 84] which could explain both the alpha decay and cluster decay under the same footing. Based on the quantum mechanical tunnelling process [85], the microscopic theories could express the decay constant $\lambda$ as a product of three model dependent parameters: frequency of assaults on the barrier in a time unit $\nu$, the cluster pre-formation probability at the nuclear surface $S$ (equal to the penetrability of the internal part of the barrier in a fission theory [83, 84]), and the probability of penetration through the external Coulomb barrier $P_S$ and the relation is given as,

$$\lambda = \ln 2 / T = \nu S P_S \quad (22)$$

By using the decimal logarithm,

$$\log_{10} T(s) = -\log_{10} P_S - \log_{10} S + [\log_{10}(\ln 2) - \log_{10} \nu] \quad (23)$$

On assuming $\nu$ to be a constant and $S$ to be depending only on the mass number of the emitted particle $A_e$ the universal formula [84] was developed and using a fit with experimental data for $\alpha$ decay, the corresponding numerical values [84] obtained were, $S_\alpha = 0.0143153$, $\nu = 10^{22.01}\text{s}^{-1}$. The decimal logarithm of the pre-formation factor is given as,

$$\log_{10} S = -0.598(A_e - 1) \quad (24)$$

and the additive constant for an even-even nucleus is,

$$c_{ee} = [-\log_{10} \nu + \log_{10}(\ln 2)] = -22.16917 \quad (25)$$

The penetrability through an external Coulomb barrier, having separation distance at the touching configuration $R_a = R_t = R_d + R_e$ as the first turning point and the second turning point defined by $e^2 Z_d Z_e / R_b = Q$, may be found analytically as

$$-\log_{10} P_S = 0.22873(\mu_A Z_d Z_e R_b)^{1/2} \times [\arccos\sqrt{r} - \sqrt{r(1-r)}] \quad (26)$$

where $r = R_t / R_b$, $R_t = 1.2249(A_d^{1/3} + A_e^{1/3})$ and $R_b = 1.43998 Z_d Z_e / Q$.

The released energy $Q$ is evaluated using the mass tables [72-75] and the liquid-drop-model radius constant $r_0 = 1.2249$ fm.

### 3.2 Empirical relations for spontaneous fission half lives

Similar to alpha decay, spontaneous fission is also a limiting factor that determines the stability of newly synthesized superheavy nuclei. On comparing the alpha decay half lives with the spontaneous fission half lives, the mode of decay of the isotopes can be predicted, and as the isotopes with shorter $\alpha$ decay half lives than the spontaneous fission half lives survive fission, those isotopes may be detected through α decay in the laboratory. In the present manuscript, the spontaneous fission half lives of all the isotopes under consideration have been evaluated using the semi-empirical relation of Xu et al. [20].

### 3.2.1 The semi-empirical relation of Xu et al.,

Recently Xu et al., [20] introduced a new approach for the spontaneous fission half-lives by using the parabolic potential approximation, taking into account the most important nuclear structure effects namely the strong interaction, the Coulomb interaction and the isospin effect. The authors thus derived a new formula from the parabolic potential approximation and systematically calculated the spontaneous fission half lives of nuclei from $^{232}$Th to $^{286}$114. The new expression for spontaneous fission half lives, originally made to fit the even-even nuclei, is given as,

$$T_{1/2} = \frac{\ln 2}{n.P_{SF}} = \exp\left\{2\pi\left[c_0 + c_1 A + c_2 Z^2 + c_3 Z^4 + c_4(N-Z)^2 - (0.13323\frac{Z^2}{A^{1/3}} - 11.64)\right]\right\} \quad (27)$$

where $n$ is the frequency factor, chosen as a constant in calculations and $P_{SF}$ is the penetration probability through the Coulomb barrier, the dominant factor in determining the spontaneous fission half lives. The values of the parameters are $c_0$ = -195.09227, $c_1$ = 3.10156, $c_2$ = -0.04386, $c_3$ = 1.4030x10$^{-6}$ and $c_4$ = -0.03199.

### 3.3 Proton separation energies

To identify the proton emitters in $^{306-339}$128, $^{288-339}$126, $^{284-339}$124 and $^{280-339}$122 superheavy nuclei, the one-proton and the two-proton separation energies [86] of all the isotopes under study were evaluated using the relations given as,

$$S(p) = -\Delta M(A,Z) + \Delta M(A-1, Z-1) + \Delta M_H = -Q(\gamma, p) \quad (28)$$

$$S(2p) = -\Delta M(A,Z) + \Delta M(A-2, Z-2) + 2\Delta M_H = -Q(\gamma, 2p) \quad (29)$$

where the terms $S(p)$ and $S(2p)$ are the one-proton separation energy and the two-proton separation energy of the nuclei, $\Delta M(A,Z)$, $\Delta M_H$, $\Delta M(A-1, Z-1)$ and $\Delta M(A-2, Z-2)$ represents respectively the mass excess of the parent nuclei, the mass excess of the proton, the mass excess of the daughter nuclei produced during the one-proton radioactivity and the mass excess of the daughter nuclei produced during the two-proton radioactivity. The $Q$ values for the one-proton radioactivity and two-proton radioactivity are given as $Q(\gamma, p)$ and $Q(\gamma, 2p)$ respectively.

### 3.4 α decay chains of Z = 128

To identify the modes of decay of the isotopes of Z = 128 within the range 306 ≤ A ≤ 339, the proton separation energy, the alpha decay half lives and the spontaneous fission half lives have been evaluated using the formalisms described above.

An evaluation of the proton separation energies for the isotopes of Z = 128 within the range 306 ≤ A ≤ 339 revealed that, the one-proton separation energy $S(p)$ is negative for the $^{306-315}$128 isotopes and the two-proton separation energy $S(2p)$ is negative for the isotopes $^{306-318}$128. Thus, the observations made it clear that all those isotopes within the range 306 ≤ A ≤ 318 lie outside the proton drip line and thus may easily decay through proton emission.

In order to find the decay modes of isotopes within the range 319 ≤ A ≤ 339, the alpha decay halflives and spontaneous fission halflives are calculated and by comparing the alpha decay half lives with the spontaneous fission half lives we have predicted the isotopes which are decaying through alpha emission. By comparing the alpha decay halflives with the spontaneous fission halflives, we could observe 6 consistent α chains from the nuclei $^{319, 320}$128, 5α chains consistently from the nuclei $^{321-324}$128, 4 consistent α chains from the nuclei $^{325,326}$128, 3α chains consistently from the nuclei $^{327-330}$128, 2 consistent α chains from $^{331,332}$128 and 1α chain from $^{333-335}$128. It is to be noted that in the case of $^{332}$128, the daughter nuclei $^{328}$126 lie outside the proton drip line and thus may easily decay through proton emission. So 1α chain can be predicted from the isotope $^{332}$128. It is seen that no isotopes with A ≥ 336 will survive fission and thus decay through spontaneous fission. Even though the nuclei $^{319,320,325-327}$128 survive fission with long alpha decay chains, these isotopes could not be predicted to be synthesised in the laboratories as the decay half lives of these parent nuclei are much below the present experimental limit (of the order of milliseconds). This underlines the fact that only those isotopes of Z = 128 within the range 321 ≤ A ≤ 324 and 328 ≤ A ≤ 335 are theoretically predictable to detect through alpha decay in laboratory.

The predictions are shown in figures 1-6, where we have plotted the logarithm of the decay half lives, $\log_{10}(T_{1/2})$, against the mass number of the parent nuclei in the corresponding alpha decay. In these figures, the plots for the decay half lives evaluated using both CPPM (considering both the parent and daughter nuclei to be spherical) and CPPMDN (which include the ground state quadrupole ($\beta_2$) and hexadecapole ($\beta_4$) deformations of both the parent and the daughter nuclei) have been shown as closed magenta circles and closed red triangles respectively. The figures show that the α half lives decreases with the inclusion of the deformation values. For a theoretical comparison, the decay half lives evaluated using the VSS formula, the UNIV and the analytical formulae of Royer have been plotted in these figures, and it can be seen that these values matches well with our theoretical calculations. The spontaneous fission half lives evaluated using the semi-empirical formulae of Xu et al., are also shown in order to predict the decay mode of the isotopes. It was observed that, most of the spontaneous fission half lives evaluated with the semi-empirical relation of Xu et al., were in good agreement with the experimental spontaneous fission half lives [87].

In Tables 1 and 2, we have highlighted our predictions on the mode of decay of unknown isotopes of Z = 128 in the range 321 ≤ A ≤ 324 and 328 ≤ A ≤ 335. In addition to the alpha decay half lives calculated using the CPPMDN, the values evaluated within CPPM, VSS formula, the analytical formulae of Royer and the UNIV have also been shown in these tables. The isotopes under study and the corresponding alpha decay products have been arranged in column 1. In column 2, the respective $Q$ values evaluated using eqn. (15) have been provided. In column 3, the spontaneous fission half lives of the corresponding isotopes evaluated using the phenomenological formula of Xu et al., have been given. The columns 4 and 5 represent respectively the calculations on the alpha half lives within both our formalisms CPPM and CPPMDN. The decay half lives calculated using the semi-empirical VSS formula, the universal curves (UNIV) and the analytical formulae of Royer have been given in column 6, column 7 and column 8 respectively. In column 9, we have shown the mode of decay of the isotopes under study. Thus, our study on the alpha decay and spontaneous fission of the isotopes of Z = 128 helps in identifying those parent and daughter nuclei that will survive fission.

**3.5 α decay chains of Z = 126**

The mode of decay and the alpha decay properties of the isotopes of Z = 126 within the range $288 \leq A \leq 339$ have been studied.

To study the behaviour of $^{288-339}126$ SHN against proton decay, we have evaluated the proton separation energies for these isotopes using the eqns. (29) and (30). It was observed that the one-proton separation energy $S(p)$ is negative for the $^{296,298-307,328}126$ isotopes and the two-proton separation energy $S(2p)$ is negative for the even-even isotopes within the range $288 \leq A \leq 298$ and also for the $^{299-312}126$ isotopes. These observations made it clear that all those even-even isotopes within the range $288 \leq A \leq 298$, isotopes within the range $299 \leq A \leq 312$ and A= 328 lie outside the proton drip line and thus may easily decay through proton emission.

By comparing the alpha decay half lives with the spontaneous fission half lives for the isotopes within the range $313 \leq A \leq 327$ and $329 \leq A \leq 339$ we could observe 6 consistent α chains from the nuclei $^{313}126$, 5α chains consistently from the nuclei $^{314-316}126$, 4 consistent α chains from the nuclei $^{317-320}126$, 3α chains consistently from the nuclei $^{321,322}126$, 2 consistent α chains from the nuclei $^{323-326}126$ and 1α chain from $^{327}126$. The isotopes within the range $329 \leq A \leq 339$ would not survive fission and hence decay through spontaneous fission. It is to be noted that, even though the nuclei $^{313-317,321,322}126$ survive fission with long alpha decay chains, these isotopes could not be predicted to be synthesised in the laboratories as the decay half lives of these parent nuclei are too short which span the order $10^{-11}$s to $10^{-7}$s. This underlines the fact that only those isotopes $^{318-320, 323-327}126$ are theoretically predictable to be synthesized and detected in laboratory through alpha decay.

As the isotopes of Z = 126 within the range $318 \leq A \leq 320$ and $324 \leq A \leq 327$ comes as the immediate daughter nuclei of $^{322-324}128$ and $^{328-331}128$ SHN, the entire study on the alpha decay half lives and the spontaneous fission half lives evaluated using the formalisms described above, can be seen from the figures 1-4. The comparison of the alpha decay half lives with the spontaneous fission half lives for the isotopes $^{323}126$ is given in figure 7.

Through these predictions, we could emphasise the fact that the isotopes $^{318-320,323-327}126$ can be synthesised and detected experimentally via alpha decay. So in Tables 1, 2 and 3, we have highlighted our predictions on the mode of decay of these isotopes of Z = 126. The predictions on the mode of decay of $^{318-320}126$, the immediate daughter of $^{322-334}128$ can be obtained from the Table 1. Similarly the decay modes of $^{324-327}126$, the daughter nuclei of $^{328-331}128$ are described in Table 2. The mode of decay of $^{323}126$ is given in Table 3. We thus hope these predictions to provide a new vision for the future experiments on SHN.

**3.6 α decay chains of Z = 124**

The proton separation energy, alpha decay half lives and the spontaneous fission half lives of the isotopes of Z = 124 SHN within the range $284 \leq A \leq 339$ have been evaluated.

The behaviour of $^{284-339}124$ SHN against the proton decay revealed that, the one-proton separation energy $S(p)$ is negative for the even-even $^{288,290}124$ isotopes and also for the $^{291-299}124$ isotopes. A similar calculation on the two-proton separation energy revealed $S(2p)$ to be negative for the even-even isotopes $^{284,286,288,290}124$ and for the $^{291-304}124$ isotopes. Thus, the observations made it clear that all the even-even isotopes within the range 284 ≤ A ≤ 290 and the isotopes within the range 291 ≤ A ≤ 304 lie outside the proton drip line and thus may easily decay through proton emission.

By comparing the alpha decay half lives with the spontaneous fission half lives within the range 305 ≤ A ≤ 339, we could observe 8 consistent α chains from the nuclei $^{305}$124, 6α chains consistently from the nuclei $^{306}$124, 5 consistent α chains from the nuclei $^{307-309}$124, 4α chains consistently from the nuclei $^{310-312}$124, 3 consistent α chains from the nuclei $^{313-316}$124, 2α chains consistently from the nuclei $^{317,318}$124 and 1α chain from the nuclei $^{319-322}$124. Any of the isotopes within the range 323 ≤ A ≤ 339 will not survive fission and hence decay through spontaneous fission. Even though the isotopes $^{309-314}$124 decay with long alpha chains, as mentioned earlier, these isotopes could not be predicted to be synthesised in the laboratories as the alpha decay half lives of these parent nuclei are too short spanning the order $10^{-15}$s to $10^{-7}$s. This underlines the fact that only those isotopes $^{305-308,315-322}$124 are theoretically predictable to be synthesized and detected trough alpha decay in laboratories.

The predictions on the isotopes of Z = 124 can be seen from figures 2-4 and 7-10. Figures 2-4 can be used to explain the decay properties of $^{315,316,320-322}$124, since these are the corresponding daughter nuclei of $^{323,324,328-330}$128. Similarly figure 7 shows the predictions of $^{319}$124, as it is the immediate daughter nuclei of $^{323}$126. Figures 8-10 shows the decay properties of $^{305-308,317,318}$124 by comparing the alpha decay half lives with the corresponding spontaneous fission half lives.

The highlights on the predictions on the decay modes of certain unknown isotopes of Z = 124 has been presented in Tables 1-4. The prediction on the isotopes $^{315, 316}$124 can be obtained from Table 1 since they are the corresponding daughter nuclei of $^{323, 324}$128. Similarly the decay modes of $^{320-322}$124 can be seen from Table 2 as they are the corresponding daughter nuclei of $^{328-330}$128. Tables 3 and 4 show the modes of decay of $^{319,305-307}$124 and $^{308,317,318}$124 respectively. Since the isotope $^{319}$124 is the immediate daughter of $^{323}$126, the mode of decay of $^{319}$124 can be obtained from the decay chain of $^{323}$126. Hence we hope our predictions on these unknown nuclei to provide a new vision for the future experiments on SHN and thus, these isotopes to be synthesized and detected experimentally via alpha decay in the laboratories.

### 3.7 α decay chains of Z = 122

Through a similar study as performed for Z = 128, Z = 126 and Z = 124, as mentioned in the above sections, the alpha decay properties and the mode of decay of all the available isotopes of Z = 122, within the range 280 ≤ A ≤ 339, has been described in the present section.

The proton decay study of these isotopes revealed that, the one-proton separation energy $S(p)$ is negative for $^{282,284-293}$122 isotopes and the two-proton separation energy $S(2p)$ is negative for the $^{280,282,284-297}$122 isotopes. Thus, the observations made it clear that the isotopes $^{280,282}$122 and the isotopes within the range 284 ≤ A ≤ 297 lie outside the proton drip line and thus may easily decay through proton emission.

The alpha decay studies of various isotopes of Z = 122 has been performed and it is seen that 7 consistent α chains from the nuclei $^{298-301}$122, 5α chains consistently from the nuclei $^{302}$122, 4 consistent α chains from the nuclei $^{303-305}$122, 3α chains consistently from the nuclei $^{306-308}$122, 2 consistent α chains from the nuclei $^{309-312}$122, 1α chain from the nuclei $^{313,314}$122. Isotopes with A ≥ 315 will not survive fission and hence decays through spontaneous fission. Even though the present study reveals that the isotopes $^{308-310}$122 decay with long alpha chains, as mentioned earlier, these isotopes could not be predicted to be synthesised experimentally as the alpha decay half is too short to be synthesised. This underlines the fact that only those isotopes $^{298-307,311-314}$122 are theoretically predictable to exhibit alpha decay.

The figures 2 and 8-13 give the predictions on the isotopes of Z = 122. Figure 2 can be used to explain the decay properties and hence predict the mode of decay of the isotopes $^{311, 312}$122, as these isotopes of Z = 122 come as the corresponding daughter nuclei of $^{323-324}$128. The plot for the isotopes $^{301,302}$122, $^{303,304}$122, $^{313,314}$122 can be seen in figures 8, 9 and 10, since these isotopes are the daughter nuclei of $^{305, 306}$124, $^{307, 308}$124 and $^{317, 318}$124. The predictions on the isotopes $^{298-300, 305-307}$122 are shown in Figures 11-13. Thus, our study on the alpha decay properties and mode of decay of heavy isotopes of Z = 122 reveals that the isotopes within the range 298 ≤ A ≤ 307 and 311 ≤ A ≤ 314 can be synthesized and detected in laboratory through alpha decay.

In Tables 4, 5 and 6, we have depicted the highlights on the predictions on the decay modes of certain unknown isotopes of Z = 122 namely $^{298}$122, $^{299,300,305}$122 and $^{306,307}$122. Our predictions on these isotopes reveal the fact that these nuclei have large alpha decay half lives, probable to be synthesised experimentally. It is to be noted that the isotopes $^{301-304, 311-314}$122 also have alpha decay half lives within the experimental limits and these predictions can be obtained from Table 1, 3 and 4. Hence, we hope that these predictions to provide a new vision for the future experiments on SHN.

For a theoretical comparison of the calculated values of spontaneous fission half lives using different models, we have calculated the average deviation of the spontaneous fission half lives of 113 isotopes of SHN. It was seen that the average deviations of spontaneous fission half lives calculated by the semi empirical formula of Xu et al., [20] with respect to the values calculated by using the formalisms of Warda et al., [21] and Staszczak et al., [22] are almost the same. The average deviation of spontaneous fission half lives calculated by the method of Warda et al., with respect to Staszczak et al., is smaller than the former cases.

# 4 Conclusion

The alpha decay properties of the isotopes of the superheavy nuclei with Z = 128, within the range 306 ≤ A ≤ 339, have been studied and thereby the mode of decay of these isotopes have been predicted extensively within CPPMDN. The decay properties of the isotopes of the daughter nuclei of Z = 128, namely Z = 126, Z = 124 and Z = 122 have also been studied in detail. A theoretical comparison of our calculations has been performed with the values evaluated using other formalisms and it should be noted that our values matches well with the values evaluated using these theoretical models. The manner in which the isotopes of these SHN would behave against the proton decay has also been considered for the present study. In order to predict the mode of decay of isotopes, the spontaneous fission half lives of the respective nuclides have been calculated. Our study on the spontaneous fission half lives and the alpha decay half lives thus highlights the range of isotopes which survive fission and thus decay through alpha emission. The significant observation on the mode of decay of $^{321-324,328-335}$128, $^{318-320,323-327}$126, $^{305-308,315-322}$124 and $^{298-307,311-314}$122 may be anticipated to be of great help for the future experimental studies to synthesize the isotopes around Z = 126.

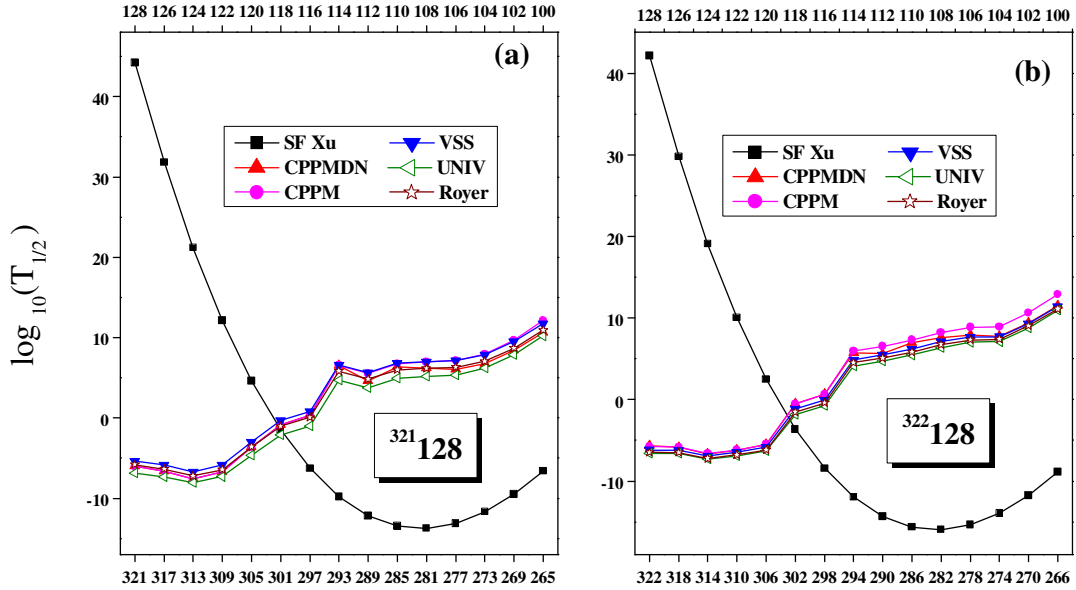

Fig 1: (Color online) Plot for the comparison of the calculated α decay half lives with the corresponding spontaneous fission half lives of the isotopes $^{321, 322}$128 and their decay products.

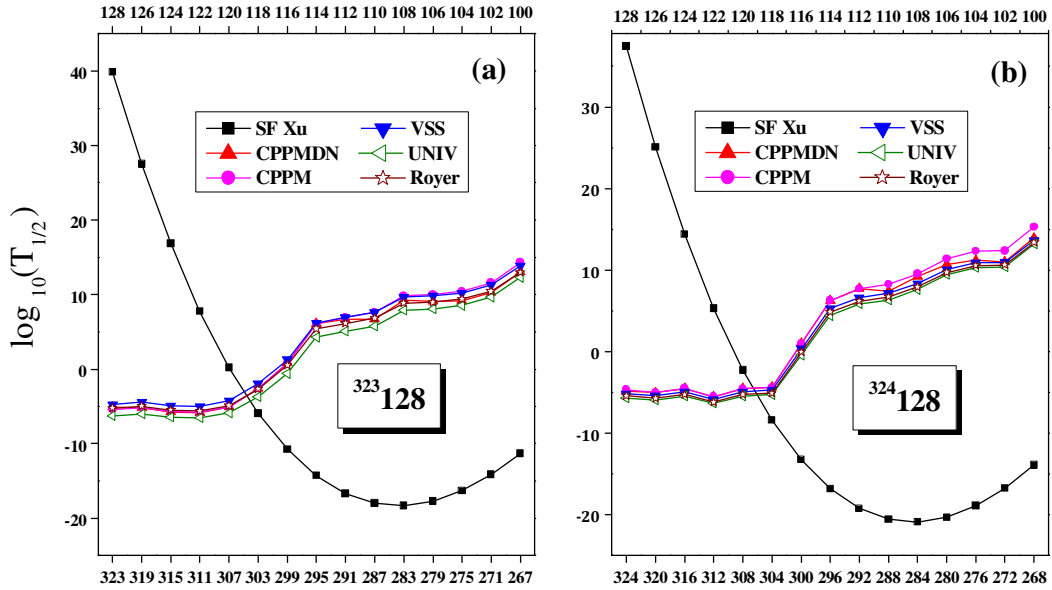

Fig 2: (Color online) Plot for the comparison of the calculated α decay half lives with the corresponding spontaneous fission half lives of the isotopes $^{323, 324}$128 and their decay products.

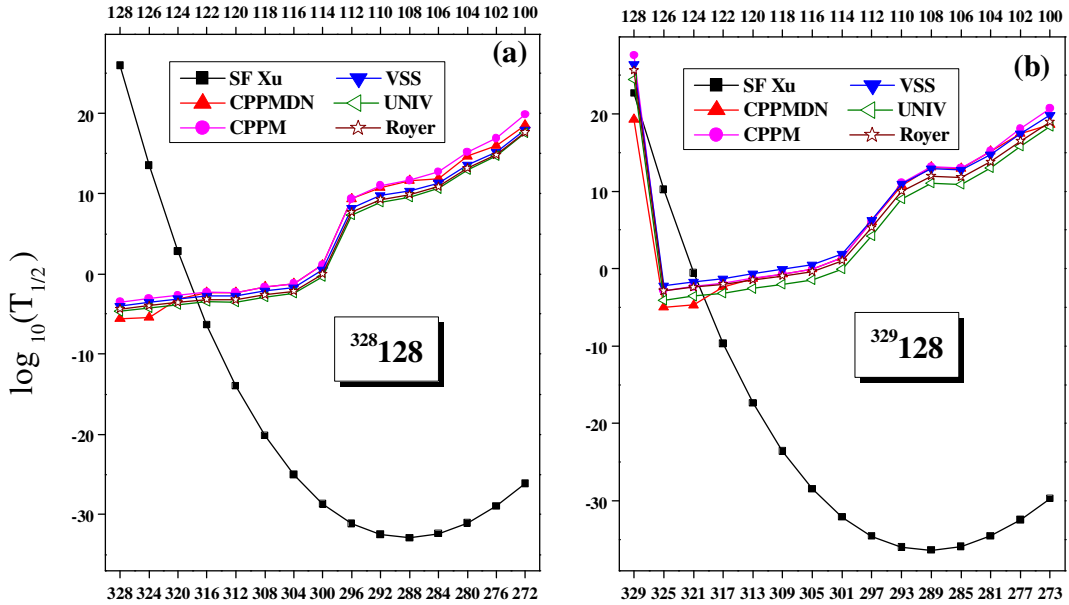

Fig 3: (Color online) Plot for the comparison of the calculated α decay half lives with the corresponding spontaneous fission half lives of the isotopes [328, 329]128 and their decay products.

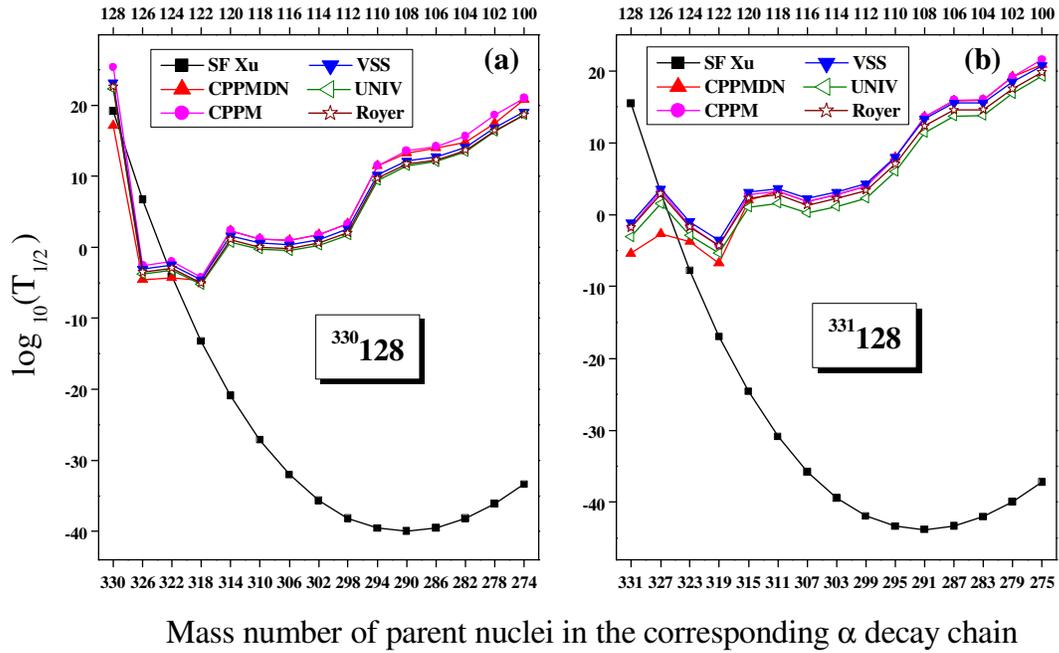

Fig 4: (Color online) Plot for the comparison of the calculated α decay half lives with the corresponding spontaneous fission half lives of the isotopes $^{330,\,331}$128 and their decay products.

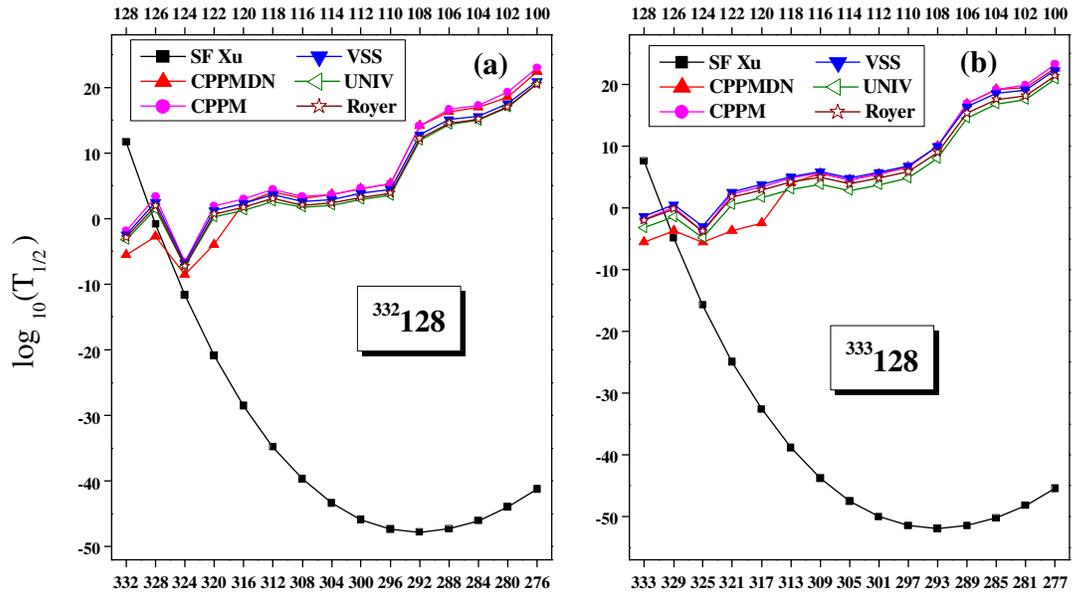

Fig 5: (Color online) Plot for the comparison of the calculated α decay half lives with the corresponding spontaneous fission half lives of the isotopes $^{332,\,333}$128 and their decay products.

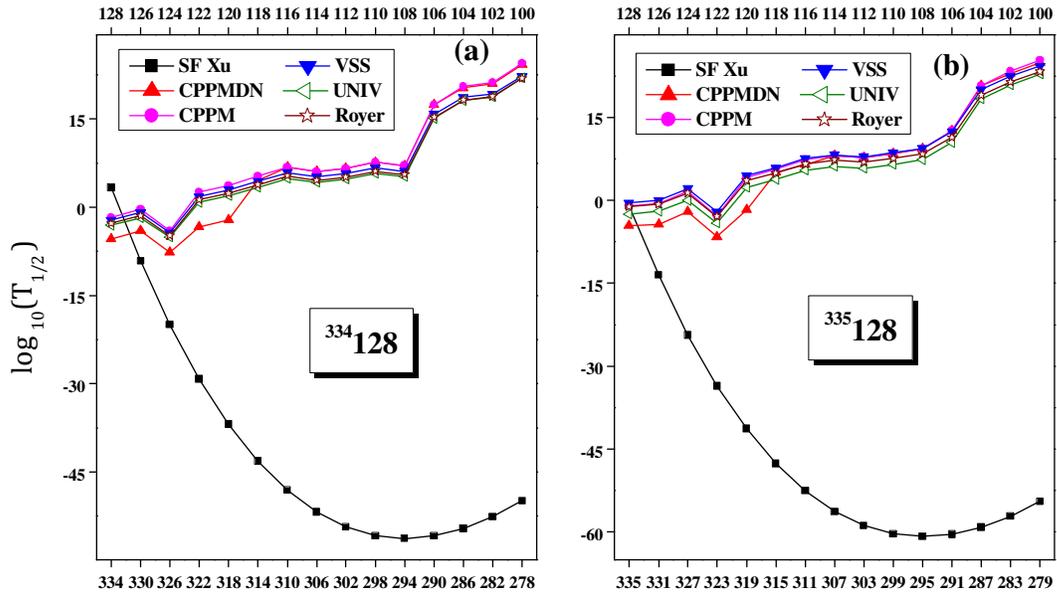

Fig 6: (Color online) Plot for the comparison of the calculated α decay half lives with the corresponding spontaneous fission half lives of the isotopes $^{334,\,335}$128 and their decay products.

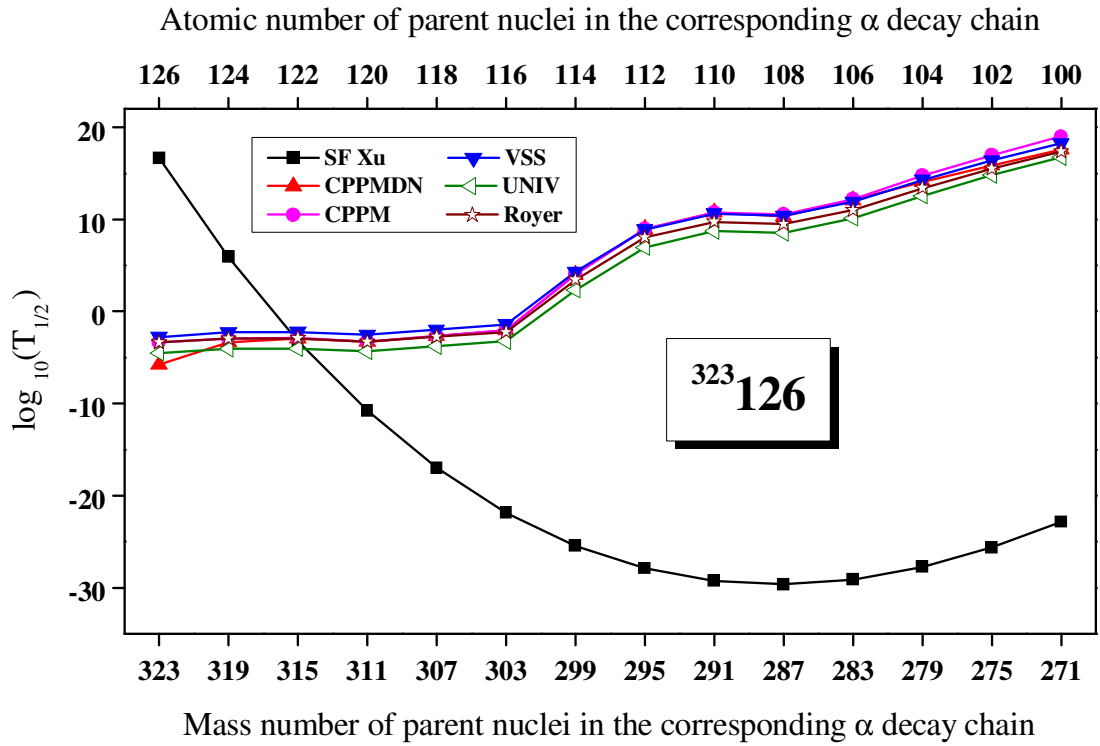

Fig 7: (Color online) Plot for the comparison of the calculated α decay half lives with the corresponding spontaneous fission half lives of the isotopes $^{323}126$ and their decay products.

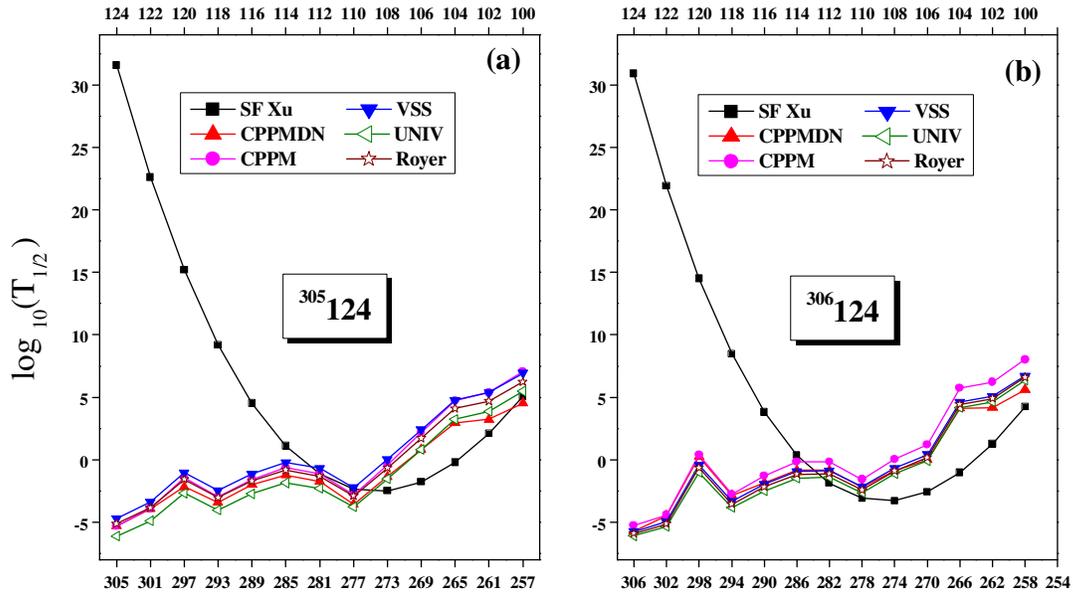

Fig 8: (Color online) Plot for the comparison of the calculated α decay half lives with the corresponding spontaneous fission half lives of the isotopes $^{305,\,306}$124 and their decay products.

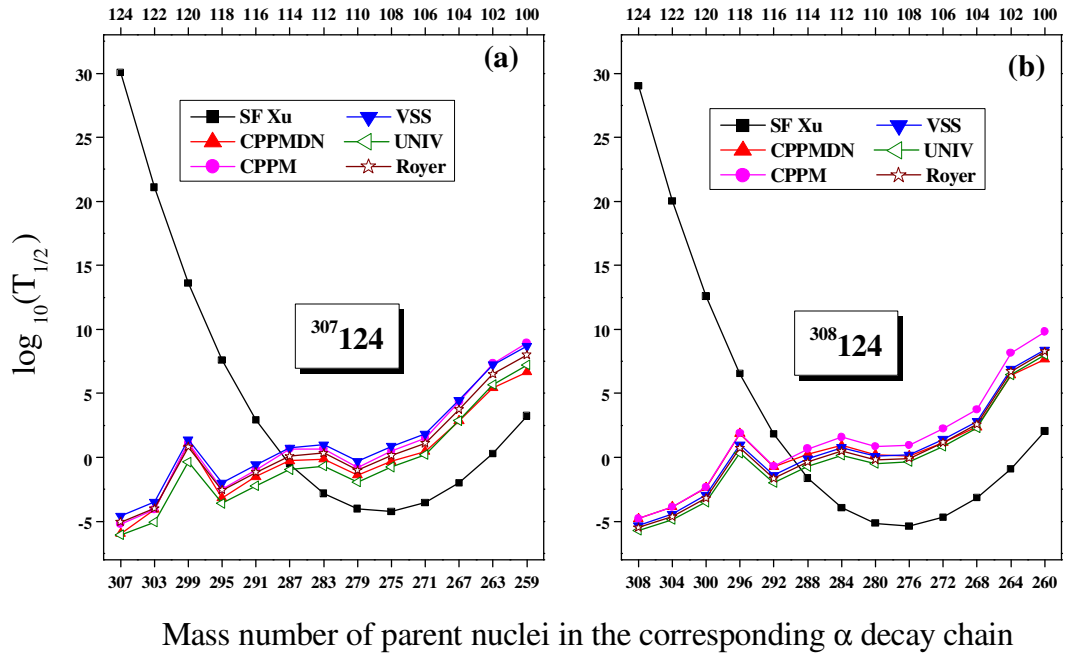

Fig 9: (Color online) Plot for the comparison of the calculated α decay half lives with the corresponding spontaneous fission half lives of the isotopes $^{307,\,308}$124 and their decay products.

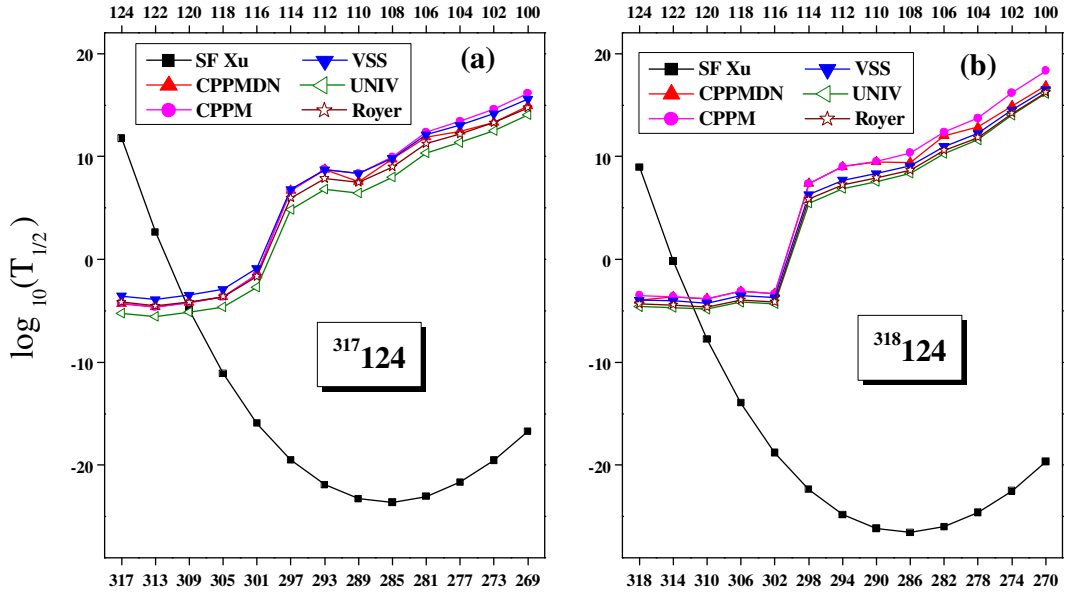

Fig 10: (Color online) Plot for the comparison of the calculated α decay half lives with the corresponding spontaneous fission half lives of isotope $^{317,\,318}$124 and its decay products.

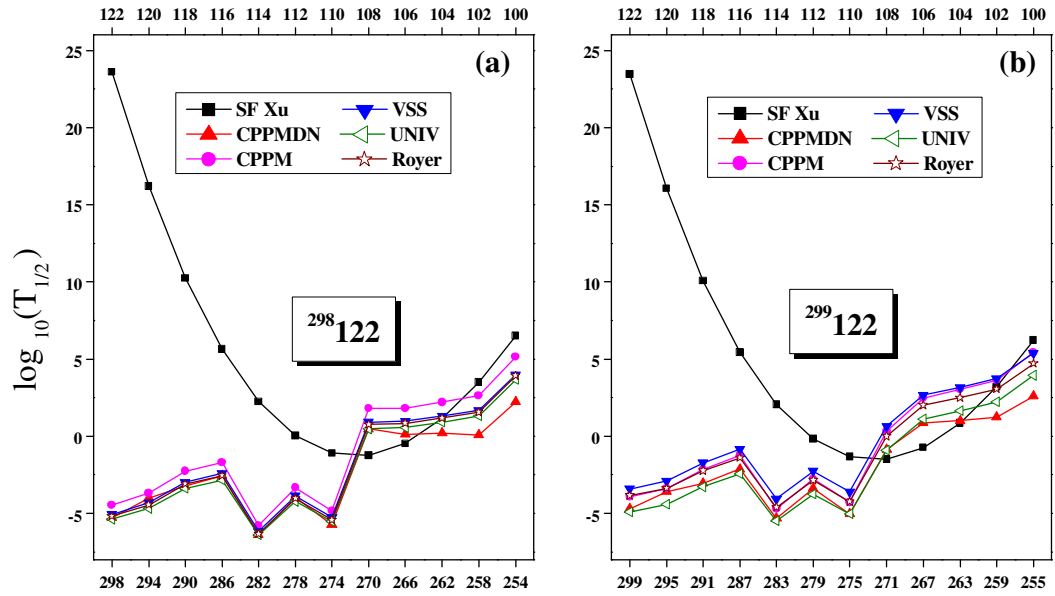

Fig 11: (Color online) Plot for the comparison of the calculated α decay half lives with the corresponding spontaneous fission half lives of the isotopes $^{298, 299}$122 and their decay products.

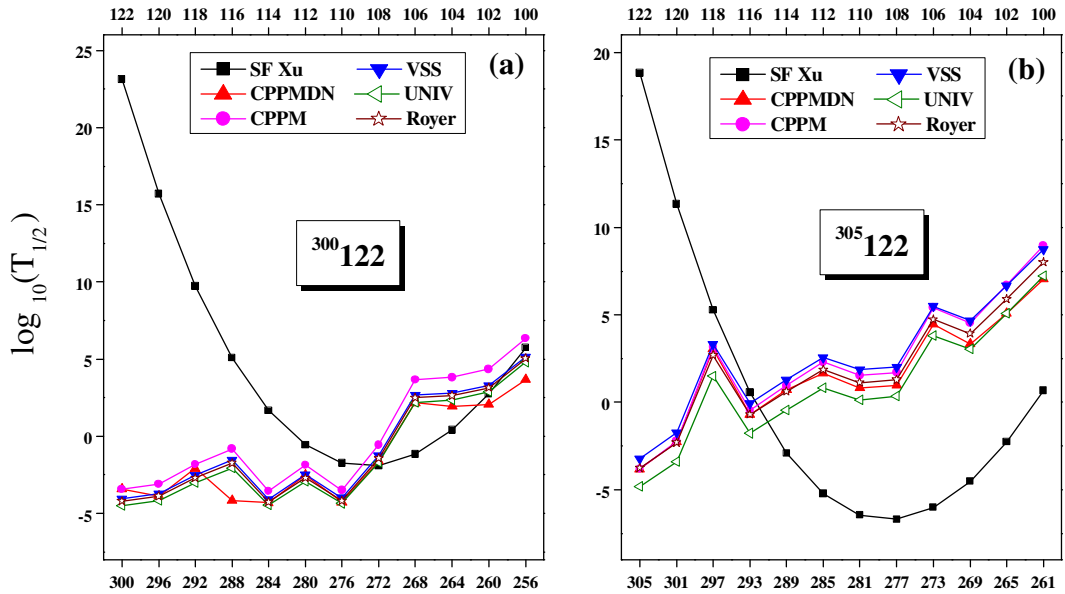

Fig 12: (Color online) Plot for the comparison of the calculated α decay half lives with the corresponding spontaneous fission half lives of the isotopes $^{300,\,305}$122 and their decay products.

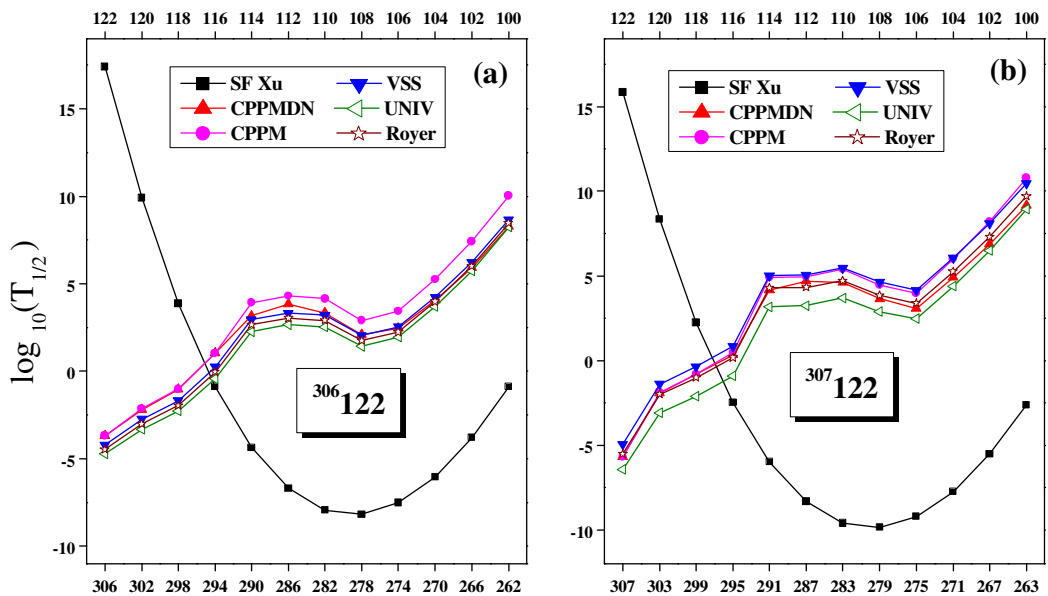

Fig 13: (Color online) Plot for the comparison of the calculated α decay half lives with the corresponding spontaneous fission half lives of the isotopes $^{306,\,307}$122 and their decay products.

Table 1: Predictions on the mode of decay of $^{321-324}128$ superheavy nuclei and their decay products is given by comparing the alpha half lives and their corresponding spontaneous fission half lives. The calculations are done for zero angular momentum transfers.

| Parent nuclei | $Q_\alpha$ (cal) (MeV) | $T_{SF}$ (s) | $T_{1/2}^\alpha$ (s) | | | | | Mode of Decay |
|---|---|---|---|---|---|---|---|---|
| | | | CPPM | CPPMDN | VSS | UNIV | Royer | |
| $^{321}128$ | 14.653 | $1.626\times10^{44}$ | $9.239\times10^{-7}$ | $1.085\times10^{-6}$ | $4.458\times10^{-6}$ | $1.479\times10^{-7}$ | $1.569\times10^{-6}$ | α1 |
| $^{317}126$ | 14.682 | $7.213\times10^{31}$ | $2.332\times10^{-7}$ | $2.263\times10^{-7}$ | $1.364\times10^{-6}$ | $4.880\times10^{-8}$ | $4.325\times10^{-7}$ | α2 |
| $^{313}124$ | 14.870 | $1.602\times10^{21}$ | $2.979\times10^{-8}$ | $2.755\times10^{-8}$ | $2.210\times10^{-7}$ | $9.100\times10^{-9}$ | $6.356\times10^{-8}$ | α3 |
| $^{309}122$ | 14.148 | $1.437\times10^{12}$ | $2.067\times10^{-7}$ | $2.058\times10^{-7}$ | $1.430\times10^{-6}$ | $4.933\times10^{-8}$ | $3.681\times10^{-7}$ | α4 |
| $^{305}120$ | 12.447 | $4.217\times10^{4}$ | $2.434\times10^{-4}$ | $2.434\times10^{-4}$ | $1.082\times10^{-3}$ | $2.353\times10^{-5}$ | $2.468\times10^{-4}$ | α5 |
| $^{301}118$ | 11.085 | $3.284\times10^{-2}$ | $1.545\times10^{-1}$ | $1.545\times10^{-1}$ | $4.649\times10^{-1}$ | $7.677\times10^{-3}$ | $9.562\times10^{-2}$ | SF |
| $^{322}128$ | 14.503 | $1.383\times10^{42}$ | $1.764\times10^{-6}$ | $2.000\times10^{-6}$ | $6.572\times10^{-7}$ | $2.558\times10^{-7}$ | $3.676\times10^{-7}$ | α1 |
| $^{318}126$ | 14.262 | $5.948\times10^{29}$ | $1.510\times10^{-6}$ | $1.455\times10^{-6}$ | $6.230\times10^{-7}$ | $2.382\times10^{-7}$ | $3.283\times10^{-7}$ | α2 |
| $^{314}124$ | 14.390 | $1.281\times10^{19}$ | $2.384\times10^{-7}$ | $2.200\times10^{-7}$ | $1.216\times10^{-7}$ | $5.217\times10^{-8}$ | $6.114\times10^{-8}$ | α3 |
| $^{310}122$ | 13.888 | $1.115\times10^{10}$ | $6.612\times10^{-7}$ | $6.576\times10^{-7}$ | $3.387\times10^{-7}$ | $1.320\times10^{-7}$ | $1.614\times10^{-7}$ | α4 |
| $^{306}120$ | 13.287 | $3.171\times10^{2}$ | $3.150\times10^{-6}$ | $3.150\times10^{-6}$ | $1.547\times10^{-6}$ | $5.359\times10^{-7}$ | $7.028\times10^{-7}$ | α5 |
| $^{302}118$ | 10.975 | $2.394\times10^{-4}$ | $2.991\times10^{-1}$ | $2.991\times10^{-1}$ | $6.957\times10^{-2}$ | $1.381\times10^{-2}$ | $2.985\times10^{-2}$ | SF |
| $^{323}128$ | 14.323 | $7.798\times10^{39}$ | $3.915\times10^{-6}$ | $9.023\times10^{-6}$ | $1.820\times10^{-5}$ | $5.040\times10^{-7}$ | $5.880\times10^{-6}$ | α1 |
| $^{319}126$ | 13.912 | $3.253\times10^{27}$ | $7.581\times10^{-6}$ | $6.447\times10^{-6}$ | $3.681\times10^{-5}$ | $9.468\times10^{-7}$ | $1.060\times10^{-5}$ | α2 |
| $^{315}124$ | 13.900 | $6.797\times10^{16}$ | $2.199\times10^{-6}$ | $1.842\times10^{-6}$ | $1.264\times10^{-5}$ | $3.466\times10^{-7}$ | $3.295\times10^{-6}$ | α3 |
| $^{311}122$ | 13.698 | $5.735\times10^{7}$ | $1.564\times10^{-6}$ | $1.412\times10^{-6}$ | $9.923\times10^{-6}$ | $2.744\times10^{-7}$ | $2.344\times10^{-6}$ | α4 |
| $^{307}120$ | 13.067 | $1.582\times10^{0}$ | $9.062\times10^{-6}$ | $8.402\times10^{-6}$ | $5.433\times10^{-5}$ | $1.326\times10^{-6}$ | $1.160\times10^{-5}$ | α5 |
| $^{303}118$ | 11.755 | $1.157\times10^{-6}$ | $2.570\times10^{-3}$ | $2.457\times10^{-3}$ | $1.092\times10^{-2}$ | $1.978\times10^{-4}$ | $2.102\times10^{-3}$ | SF |
| $^{324}128$ | 13.963 | $2.917\times10^{37}$ | $2.091\times10^{-5}$ | $1.313\times10^{-5}$ | $6.992\times10^{-6}$ | $2.123\times10^{-6}$ | $3.557\times10^{-6}$ | α1 |
| $^{320}126$ | 13.832 | $1.180\times10^{25}$ | $1.076\times10^{-5}$ | $9.158\times10^{-6}$ | $4.120\times10^{-6}$ | $1.275\times10^{-6}$ | $1.982\times10^{-6}$ | α2 |
| $^{316}124$ | 13.339 | $2.392\times10^{14}$ | $3.271\times10^{-5}$ | $2.864\times10^{-5}$ | $1.246\times10^{-5}$ | $3.524\times10^{-6}$ | $5.659\times10^{-6}$ | α3 |
| $^{312}122$ | 13.558 | $1.957\times10^{5}$ | $2.959\times10^{-6}$ | $2.862\times10^{-6}$ | $1.446\times10^{-6}$ | $4.716\times10^{-7}$ | $6.314\times10^{-7}$ | α4 |
| $^{308}120$ | 12.837 | $5.233\times10^{-3}$ | $2.820\times10^{-5}$ | $2.813\times10^{-5}$ | $1.257\times10^{-5}$ | $3.528\times10^{-6}$ | $5.229\times10^{-6}$ | α5 |
| $^{304}118$ | 12.495 | $3.713\times10^{-9}$ | $4.218\times10^{-5}$ | $4.218\times10^{-5}$ | $1.937\times10^{-5}$ | $5.325\times10^{-6}$ | $7.774\times10^{-6}$ | SF |

Table 2: Predictions on the mode of decay of $^{328-335}$128 superheavy nuclei and their decay products is given by comparing the alpha half lives and their corresponding spontaneous fission half lives. The calculations are done for zero angular momentum transfers.

| Parent nuclei | $Q_\alpha$ (cal) (MeV) | $T_{SF}$ (s) | $T_{1/2}^\alpha$ (s) | | | | | Mode of Decay |
|---|---|---|---|---|---|---|---|---|
| | | | CPPM | CPPMDN | VSS | UNIV | Royer | |
| $^{328}$128 | 13.393 | 9.424x10$^{25}$ | 3.127x10$^{-4}$ | 2.798x10$^{-6}$ | 9.892x10$^{-5}$ | 2.186x10$^{-5}$ | 4.215x10$^{-5}$ | α1 |
| $^{324}$126 | 12.942 | 3.377x10$^{13}$ | 8.574x10$^{-4}$ | 3.971x10$^{-6}$ | 2.762x10$^{-4}$ | 5.645x10$^{-5}$ | 1.106x10$^{-4}$ | α2 |
| $^{320}$124 | 12.499 | 6.056x10$^{2}$ | 2.374x10$^{-3}$ | 7.797x10$^{-4}$ | 7.646x10$^{-4}$ | 1.464x10$^{-4}$ | 2.898x10$^{-4}$ | α3 |
| $^{316}$122 | 12.098 | 4.383x10$^{-7}$ | 5.465x10$^{-3}$ | 4.878x10$^{-3}$ | 1.787x10$^{-3}$ | 3.272x10$^{-4}$ | 6.462x10$^{-4}$ | SF |
| $^{329}$128 | 5.763 | 4.531x10$^{22}$ | 3.481x10$^{27}$ | 1.752x10$^{19}$ | 2.305x10$^{26}$ | 2.633x10$^{24}$ | 3.716x10$^{25}$ | α1 |
| $^{325}$126 | 12.832 | 1.575x10$^{10}$ | 1.495x10$^{-3}$ | 9.572x10$^{-6}$ | 6.117x10$^{-3}$ | 9.150x10$^{-5}$ | 1.361x10$^{-3}$ | α2 |
| $^{321}$124 | 12.360 | 2.739x10$^{-1}$ | 4.975x10$^{-3}$ | 1.931x10$^{-5}$ | 2.021x10$^{-2}$ | 2.811x10$^{-4}$ | 4.035x10$^{-3}$ | α3 |
| $^{317}$122 | 11.938 | 1.923x10$^{-10}$ | 1.328x10$^{-2}$ | 4.372x10$^{-3}$ | 5.392x10$^{-2}$ | 7.144x10$^{-4}$ | 9.743x10$^{-3}$ | SF |
| $^{330}$128 | 6.053 | 1.445x10$^{19}$ | 2.194x10$^{25}$ | 1.175x10$^{17}$ | 1.562x10$^{23}$ | 1.950x10$^{22}$ | 4.344x10$^{22}$ | α1 |
| $^{326}$126 | 12.712 | 4.874x10$^{6}$ | 2.773x10$^{-3}$ | 2.538x10$^{-5}$ | 8.791x10$^{-4}$ | 1.568x10$^{-4}$ | 3.229x10$^{-4}$ | α2 |
| $^{322}$124 | 12.229 | 8.223x10$^{-5}$ | 1.009x10$^{-2}$ | 4.985x10$^{-5}$ | 3.139x10-3 | 5.200x10$^{-4}$ | 1.091x10$^{-3}$ | α3 |
| $^{318}$122 | 12.938 | 5.598x10$^{-14}$ | 5.382x10$^{-5}$ | 2.328x10$^{-5}$ | 2.566x10-5 | 5.588x10$^{-6}$ | 8.680x10$^{-6}$ | SF |
| $^{331}$128 | 12.593 | 3.059x10$^{15}$ | 2.079x10$^{-2}$ | 4.090x10$^{-6}$ | 7.027x10$^{-2}$ | 8.673x10$^{-4}$ | 1.589x10$^{-2}$ | α1 |
| $^{327}$126 | 10.532 | 1.001x10$^{3}$ | 2.268x10$^{3}$ | 2.201x10$^{-3}$ | 4.051x10$^{3}$ | 3.270x10$^{1}$ | 7.725x10$^{2}$ | α2 |
| $^{323}$124 | 12.040 | 1.638x10$^{-8}$ | 2.900x10$^{-2}$ | 1.902x10$^{-4}$ | 1.115x10$^{-1}$ | 1.327x10$^{-3}$ | 2.047x10$^{-2}$ | SF |
| $^{332}$128 | 12.663 | 4.297x10$^{11}$ | 1.360x10$^{-2}$ | 2.700x10$^{-6}$ | 3.810x10$^{-3}$ | 5.920x10$^{-4}$ | 1.355x10$^{-3}$ | α1 |
| $^{328}$126 | 10.542 | 1.364x10$^{-1}$ | 2.041x10$^{3}$ | 1.981x10$^{-3}$ | 2.962x10$^{2}$ | 2.954x10$^{1}$ | 9.435x10$^{1}$ | *PD |
| $^{333}$128 | 12.663 | 4.006x10$^{7}$ | 1.317x10$^{-2}$ | 2.549x10$^{-6}$ | 4.870x10$^{-2}$ | 5.718x10$^{-4}$ | 1.024x10$^{-2}$ | α1 |
| $^{329}$126 | 11.662 | 1.234x10$^{-5}$ | 1.126x10$^{0}$ | 2.213x10$^{-4}$ | 3.406x10$^{0}$ | 3.244x10$^{-2}$ | 6.284x10$^{-1}$ | SF |
| $^{334}$128 | 12.573 | 2.479x10$^{3}$ | 2.110x10$^{-2}$ | 4.150x10$^{-6}$ | 6.109x10$^{-3}$ | 8.637x10$^{-4}$ | 2.000x10$^{-3}$ | α1 |
| $^{330}$126 | 11.782 | 7.406x10$^{-10}$ | 5.216x10$^{-1}$ | 1.042x10$^{-4}$ | 1.335x10$^{-1}$ | 1.615x10$^{-2}$ | 4.075x10$^{-2}$ | SF |
| $^{335}$128 | 12.293 | 1.018x10$^{-1}$ | 1.015x10$^{-1}$ | 2.709x10$^{-5}$ | 3.505x10$^{-1}$ | 3.471x10$^{-3}$ | 6.756x10$^{-2}$ | α1 |
| $^{331}$126 | 11.872 | 2.951x10$^{-14}$ | 2.927x10$^{-1}$ | 4.342x10$^{-5}$ | 1.023x10$^{0}$ | 9.563x10$^{-3}$ | 1.763x10$^{-1}$ | SF |

*PD denotes the proton decay

Table 3: Predictions on the mode of decay of $^{323}126$ and $^{305-307}124$ superheavy nuclei and their decay products is given by comparing the alpha half lives and their corresponding spontaneous fission half lives. The calculations are done for zero angular momentum transfers.

| Parent nuclei | $Q_\alpha$ (cal) (MeV) | $T_{SF}$ (s) | $T_{1/2}^\alpha$ (s) | | | | | Mode of Decay |
|---|---|---|---|---|---|---|---|---|
| | | | CPPM | CPPMDN | VSS | UNIV | Royer | |
| $^{323}126$ | 13.082 | $4.805 \times 10^{16}$ | $4.234 \times 10^{-4}$ | $1.608 \times 10^{-6}$ | $1.771 \times 10^{-3}$ | $3.056 \times 10^{-5}$ | $4.280 \times 10^{-4}$ | α1 |
| $^{319}124$ | 12.620 | $8.883 \times 10^{5}$ | $1.265 \times 10^{-3}$ | $4.431 \times 10^{-4}$ | $5.293 \times 10^{-3}$ | $8.496 \times 10^{-5}$ | $1.148 \times 10^{-3}$ | α2 |
| $^{315}122$ | 12.348 | $6.629 \times 10^{-4}$ | $1.388 \times 10^{-3}$ | $1.241 \times 10^{-3}$ | $6.176 \times 10^{-3}$ | $9.815 \times 10^{-5}$ | $1.216 \times 10^{-3}$ | SF |
| $^{305}124$ | 13.800 | $3.623 \times 10^{31}$ | $5.091 \times 10^{-6}$ | $4.806 \times 10^{-6}$ | $1.964 \times 10^{-5}$ | $7.395 \times 10^{-7}$ | $7.545 \times 10^{-6}$ | α1 |
| $^{301}122$ | 12.888 | $4.162 \times 10^{22}$ | $1.263 \times 10^{-4}$ | $1.232 \times 10^{-4}$ | $4.173 \times 10^{-4}$ | $1.275 \times 10^{-5}$ | $1.432 \times 10^{-4}$ | α2 |
| $^{297}120$ | 11.607 | $1.565 \times 10^{15}$ | $3.807 \times 10^{-2}$ | $6.598 \times 10^{-3}$ | $9.079 \times 10^{-2}$ | $2.113 \times 10^{-3}$ | $2.788 \times 10^{-2}$ | α3 |
| $^{293}118$ | 11.975 | $1.564 \times 10^{9}$ | $1.052 \times 10^{-3}$ | $3.756 \times 10^{-4}$ | $3.416 \times 10^{-3}$ | $9.420 \times 10^{-5}$ | $9.770 \times 10^{-4}$ | α4 |
| $^{289}$Lv | 11.164 | $3.384 \times 10^{4}$ | $2.773 \times 10^{-2}$ | $1.052 \times 10^{-2}$ | $7.568 \times 10^{-2}$ | $1.850 \times 10^{-3}$ | $1.989 \times 10^{-2}$ | α5 |
| $^{285}$Fl | 10.572 | $1.296 \times 10^{1}$ | $2.483 \times 10^{-1}$ | $5.569 \times 10^{-2}$ | $6.069 \times 10^{-1}$ | $1.419 \times 10^{-2}$ | $1.485 \times 10^{-1}$ | α6 |
| $^{281}$Cn | 10.521 | $7.205 \times 10^{-2}$ | $7.163 \times 10^{-2}$ | $1.814 \times 10^{-2}$ | $1.991 \times 10^{-1}$ | $5.041 \times 10^{-3}$ | $4.591 \times 10^{-2}$ | α7 |
| $^{277}$Ds | 10.889 | $4.782 \times 10^{-3}$ | $1.526 \times 10^{-3}$ | $2.780 \times 10^{-4}$ | $5.782 \times 10^{-3}$ | $1.756 \times 10^{-4}$ | $1.261 \times 10^{-3}$ | α8 |
| $^{273}$Hs | 9.778 | $3.129 \times 10^{-3}$ | $4.088 \times 10^{-1}$ | $4.503 \times 10^{-2}$ | $1.059 \times 10^{0}$ | $2.830 \times 10^{-2}$ | $2.210 \times 10^{-1}$ | SF |
| $^{306}124$ | 13.770 | $7.751 \times 10^{30}$ | $5.657 \times 10^{-6}$ | $2.038 \times 10^{-6}$ | $1.756 \times 10^{-6}$ | $8.077 \times 10^{-7}$ | $1.218 \times 10^{-6}$ | α1 |
| $^{302}122$ | 13.118 | $8.633 \times 10^{21}$ | $3.731 \times 10^{-5}$ | $3.393 \times 10^{-5}$ | $1.090 \times 10^{-5}$ | $4.398 \times 10^{-6}$ | $7.172 \times 10^{-6}$ | α2 |
| $^{298}120$ | 10.927 | $3.147 \times 10^{14}$ | $2.539 \times 10^{0}$ | $1.768 \times 10^{0}$ | $4.564 \times 10^{-3}$ | $1.334 \times 10^{-3}$ | $2.837 \times 10^{-3}$ | α3 |
| $^{294}118$ | 11.875 | $3.048 \times 10^{8}$ | $1.782 \times 10^{-3}$ | $1.374 \times 10^{-3}$ | $1.616 \times 10^{-2}$ | $4.557 \times 10^{-3}$ | $9.732 \times 10^{-3}$ | α4 |
| $^{290}$Lv | 11.054 | $6.392 \times 10^{3}$ | $5.259 \times 10^{-2}$ | $1.484 \times 10^{-2}$ | $1.349 \times 10^{-1}$ | $3.617 \times 10^{-2}$ | $7.929 \times 10^{-2}$ | α5 |
| $^{286}$Fl | 10.422 | $2.372 \times 10^{0}$ | $6.729 \times 10^{-1}$ | $1.434 \times 10^{-1}$ | $7.581 \times 10^{0}$ | $1.889 \times 10^{0}$ | $4.392 \times 10^{0}$ | α6 |
| $^{282}$Cn | 10.171 | $1.277 \times 10^{-2}$ | $6.986 \times 10^{-1}$ | $1.430 \times 10^{-1}$ | $4.953 \times 10^{-4}$ | $1.770 \times 10^{-4}$ | $2.849 \times 10^{-4}$ | SF |
| $^{307}124$ | 13.740 | $1.099 \times 10^{30}$ | $6.292 \times 10^{-6}$ | $1.054 \times 10^{-6}$ | $2.566 \times 10^{-5}$ | $8.827 \times 10^{-7}$ | $9.096 \times 10^{-6}$ | α1 |
| $^{303}122$ | 12.948 | $1.187 \times 10^{21}$ | $8.577 \times 10^{-5}$ | $8.539 \times 10^{-5}$ | $3.126 \times 10^{-4}$ | $9.044 \times 10^{-6}$ | $9.927 \times 10^{-5}$ | α2 |
| $^{299}120$ | 10.667 | $4.195 \times 10^{13}$ | $1.370 \times 10^{1}$ | $1.199 \times 10^{1}$ | $2.370 \times 10^{1}$ | $4.170 \times 10^{-1}$ | $6.586 \times 10^{0}$ | α3 |
| $^{295}118$ | 11.765 | $3.939 \times 10^{7}$ | $3.220 \times 10^{-3}$ | $6.596 \times 10^{-4}$ | $1.035 \times 10^{-2}$ | $2.512 \times 10^{-4}$ | $2.726 \times 10^{-3}$ | α4 |
| $^{291}$Lv | 10.944 | $8.005 \times 10^{2}$ | $1.008 \times 10^{-1}$ | $3.027 \times 10^{-2}$ | $2.695 \times 10^{-1}$ | $5.817 \times 10^{-3}$ | $6.523 \times 10^{-2}$ | α5 |
| $^{287}$Fl | 10.222 | $2.878 \times 10^{-1}$ | $4.355 \times 10^{0}$ | $5.260 \times 10^{-1}$ | $5.342 \times 10^{0}$ | $1.081 \times 10^{-1}$ | $1.203 \times 10^{0}$ | SF |

Table 4: Predictions on the mode of decay of $^{308,317,318}124$ and $^{298}122$ superheavy nuclei and their decay products is given by comparing the alpha half lives and their corresponding spontaneous fission half lives. The calculations are done for zero angular momentum transfers.

| Parent nuclei | $Q_\alpha$ (cal) (MeV) | $T_{SF}$ (s) | $T_{1/2}^\alpha$ (s) | | | | | Mode of Decay |
|---|---|---|---|---|---|---|---|---|
| | | | CPPM | CPPMDN | VSS | UNIV | Royer | |
| $^{308}$124 | 13.540 | 1.034x10$^{29}$ | 1.603x10$^{-5}$ | 1.599x10$^{-5}$ | 4.952x10$^{-6}$ | 1.973x10$^{-6}$ | 3.145x10$^{-6}$ | α1 |
| $^{304}$122 | 12.858 | 1.082x10$^{20}$ | 1.319x10$^{-4}$ | 1.284x10$^{-4}$ | 3.775x10$^{-5}$ | 1.312x10$^{-5}$ | 2.273x10$^{-5}$ | α2 |
| $^{300}$120 | 11.947 | 3.708x10$^{12}$ | 4.668x10$^{-3}$ | 4.065x10$^{-3}$ | 1.118x10$^{-3}$ | 3.243x10$^{-4}$ | 6.411x10$^{-4}$ | α3 |
| $^{296}$118 | 10.185 | 3.374x10$^{6}$ | 7.540x10$^{1}$ | 6.812x10$^{1}$ | 5.824x10$^{-2}$ | 1.451x10$^{-2}$ | 3.214x10$^{-2}$ | α4 |
| $^{292}$Lv | 10.834 | 6.646x10$^{1}$ | 1.951x10$^{-1}$ | 1.732x10$^{-1}$ | 6.005x10$^{-1}$ | 1.425x10$^{-1}$ | 3.237x10$^{-1}$ | α5 |
| $^{288}$Fl | 10.122 | 2.316x10$^{-2}$ | 4.583x10$^{0}$ | 1.751x10$^{0}$ | 4.615x10$^{1}$ | 1.026x10$^{1}$ | 2.453x10$^{1}$ | SF |
| $^{317}$124 | 13.220 | 5.583x10$^{11}$ | 5.768x10$^{-5}$ | 5.147x10$^{-5}$ | 2.800x10$^{-4}$ | 5.762x10$^{-6}$ | 6.656x10$^{-5}$ | α1 |
| $^{313}$122 | 13.128 | 4.430x10$^{2}$ | 2.398x10$^{-5}$ | 2.262x10$^{-5}$ | 1.330x10$^{-4}$ | 2.856x10$^{-6}$ | 2.874x10$^{-5}$ | α2 |
| $^{309}$120 | 12.677 | 1.149x10$^{-5}$ | 6.263x10$^{-5}$ | 6.231x10$^{-5}$ | 3.477x10$^{-4}$ | 7.023x10$^{-6}$ | 6.824x10$^{-5}$ | SF |
| $^{318}$124 | 12.889 | 8.646x10$^{8}$ | 3.097x10$^{-4}$ | 1.078x10$^{-4}$ | 1.076x10$^{-4}$ | 2.472x10$^{-5}$ | 4.459x10$^{-5}$ | α1 |
| $^{314}$122 | 12.678 | 6.653x10$^{-1}$ | 2.396x10$^{-4}$ | 2.200x10$^{-4}$ | 9.122x10$^{-5}$ | 2.110x10$^{-5}$ | 3.616x10$^{-5}$ | α2 |
| $^{310}$120 | 12.517 | 1.672x10$^{-8}$ | 1.413x10$^{-4}$ | 1.401x10$^{-4}$ | 5.973x10$^{-5}$ | 1.420x10$^{-5}$ | 2.278x10$^{-5}$ | SF |
| $^{298}$122 | 13.168 | 3.955x10$^{23}$ | 3.383x10$^{-5}$ | 5.207x10$^{-6}$ | 8.622x10$^{-6}$ | 4.078x10$^{-6}$ | 6.729x10$^{-6}$ | α1 |
| $^{294}$120 | 12.557 | 1.633x10$^{16}$ | 2.031x10$^{-4}$ | 1.029x10$^{-4}$ | 4.900x10$^{-5}$ | 2.089x10$^{-5}$ | 3.658x10$^{-5}$ | α2 |
| $^{290}$118 | 11.705 | 1.792x10$^{10}$ | 5.512x10$^{-3}$ | 5.766x10$^{-4}$ | 1.118x10$^{-3}$ | 4.106x10$^{-4}$ | 8.034x10$^{-4}$ | α3 |
| $^{286}$Lv | 11.234 | 4.263x10$^{5}$ | 2.032x10$^{-2}$ | 2.598x10$^{-3}$ | 3.984x10$^{-3}$ | 1.416x10$^{-3}$ | 2.792x10$^{-3}$ | α4 |
| $^{282}$Fl | 12.682 | 1.795x10$^{2}$ | 1.669x10$^{-6}$ | 4.061x10$^{-7}$ | 2.042x10$^{-4}$ | 8.528x10$^{-5}$ | 1.411x10$^{-4}$ | α5 |
| $^{278}$Cn | 11.361 | 1.098x10$^{0}$ | 4.778x10$^{-4}$ | 1.111x10$^{-4}$ | 4.459x10$^{-7}$ | 2.709x10$^{-7}$ | 3.038x10$^{-7}$ | α6 |
| $^{274}$Ds | 11.719 | 8.017x10$^{-2}$ | 1.484x10$^{-5}$ | 1.883x10$^{-6}$ | 5.635x10$^{-6}$ | 3.029x10$^{-6}$ | 3.816x10$^{-6}$ | α7 |
| $^{270}$Hs | 9.098 | 5.776x10$^{-2}$ | 6.601x10$^{1}$ | 2.905x10$^{0}$ | 8.406x10$^{0}$ | 2.908x10$^{0}$ | 5.876x10$^{0}$ | SF |

Table 5: Predictions on the mode of decay of $^{299,300,305}$122 superheavy nuclei and their decay products is given by comparing the alpha half lives and their corresponding spontaneous fission half lives. The calculations are done for zero angular momentum transfers.

| Parent nuclei | $Q_\alpha$ (cal) (MeV) | $T_{SF}$ (s) | $T_{1/2}^\alpha$ (s) | | | | | Mode of Decay |
| --- | --- | --- | --- | --- | --- | --- | --- | --- |
| | | | CPPM | CPPMDN | VSS | UNIV | Royer | |
| $^{299}$122 | 12.908 | 2.817x10$^{23}$ | 1.230x10$^{-4}$ | 1.928x10$^{-5}$ | 3.789x10$^{-4}$ | 1.252x10$^{-5}$ | 1.409x10$^{-4}$ | α1 |
| $^{295}$120 | 12.417 | 1.127x10$^{16}$ | 4.137x10$^{-4}$ | 2.657x10$^{-4}$ | 1.258x10$^{-3}$ | 3.890x10$^{-5}$ | 4.251x10$^{-4}$ | α2 |
| $^{291}$118 | 11.655 | 1.199x10$^{10}$ | 7.090x10$^{-3}$ | 9.178x10$^{-4}$ | 1.873x10$^{-2}$ | 5.121x10$^{-4}$ | 5.773x10$^{-3}$ | α3 |
| $^{287}$Lv | 11.054 | 2.764x10$^{5}$ | 5.893x10$^{-2}$ | 7.659x10$^{-3}$ | 1.421x10$^{-1}$ | 3.652x10$^{-3}$ | 4.042x10$^{-2}$ | α4 |
| $^{283}$Fl | 12.192 | 1.128x10$^{2}$ | 1.996x10$^{-5}$ | 4.754x10$^{-6}$ | 9.008x10$^{-5}$ | 3.392x10$^{-6}$ | 2.419x10$^{-5}$ | α5 |
| $^{279}$Cn | 11.141 | 6.679x10$^{-1}$ | 1.666x10$^{-3}$ | 4.077x10$^{-4}$ | 5.671x10$^{-3}$ | 1.781x10$^{-4}$ | 1.420x10$^{-3}$ | α6 |
| $^{275}$Ds | 11.479 | 4.725x10$^{-2}$ | 5.353x10$^{-5}$ | 8.923x10$^{-6}$ | 2.437x10$^{-4}$ | 9.253x10$^{-6}$ | 5.759x10$^{-5}$ | α7 |
| $^{271}$Hs | 9.558 | 3.297x10$^{-2}$ | 2.077x10$^{0}$ | 1.286x10$^{-1}$ | 4.472x10$^{0}$ | 1.243x10$^{-1}$ | 1.014x10$^{0}$ | SF |
| $^{300}$122 | 12.688 | 1.330x10$^{23}$ | 3.760x10$^{-4}$ | 1.235x10$^{-4}$ | 8.681x10$^{-5}$ | 3.317x10$^{-5}$ | 6.171x10$^{-5}$ | α1 |
| $^{296}$120 | 12.287 | 5.159x10$^{15}$ | 8.082x10$^{-4}$ | 1.300x10$^{-4}$ | 1.900x10$^{-4}$ | 6.988x10$^{-5}$ | 1.298x10$^{-4}$ | α2 |
| $^{292}$118 | 11.525 | 5.320x10$^{9}$ | 1.466x10$^{-2}$ | 8.092x10$^{-3}$ | 2.985x10$^{-3}$ | 9.735x10$^{-4}$ | 1.965x10$^{-3}$ | α3 |
| $^{288}$Lv | 10.894 | 1.188x10$^{5}$ | 1.547x10$^{-1}$ | 6.771x10$^{-5}$ | 2.828x10$^{-2}$ | 8.637x10$^{-3}$ | 1.815x10$^{-2}$ | α4 |
| $^{284}$Fl | 11.702 | 4.696x10$^{1}$ | 2.780x10$^{-4}$ | 4.959x10$^{-5}$ | 1.347x10$^{-1}$ | 4.040x10$^{-2}$ | 8.515x10$^{-2}$ | α5 |
| $^{280}$Cn | 10.791 | 2.694x10$^{-1}$ | 1.345x10$^{-2}$ | 3.084x10$^{-3}$ | 3.265x10$^{-6}$ | 1.628x10$^{-6}$ | 2.042x10$^{-6}$ | α6 |
| $^{276}$Ds | 11.159 | 1.846x10$^{-2}$ | 3.198x10$^{-4}$ | 5.449x10$^{-5}$ | 1.029x10$^{-4}$ | 4.431x10$^{-5}$ | 6.417x10$^{-5}$ | α7 |
| $^{272}$Hs | 9.838 | 1.247x10$^{-2}$ | 2.801x10$^{-1}$ | 2.424x10$^{-2}$ | 5.639x10$^{-2}$ | 2.023x10$^{-2}$ | 3.573x10$^{-2}$ | SF |
| $^{305}$122 | 12.818 | 6.544x10$^{18}$ | 1.567x10$^{-4}$ | 1.509x10$^{-4}$ | 5.861x10$^{-4}$ | 1.519x10$^{-5}$ | 1.716x10$^{-4}$ | α1 |
| $^{301}$120 | 11.897 | 2.173x10$^{11}$ | 5.993x10$^{-3}$ | 5.153x10$^{-3}$ | 1.866x10$^{-2}$ | 4.032x10$^{-4}$ | 4.922x10$^{-3}$ | α2 |
| $^{297}$118 | 9.775 | 1.917x10$^{5}$ | 1.465x10$^{3}$ | 1.116x10$^{3}$ | 2.107x10$^{3}$ | 3.159x10$^{1}$ | 4.942x10$^{2}$ | α3 |
| $^{293}$Lv | 10.744 | 3.659x10$^{0}$ | 3.355x10$^{-1}$ | 1.898x10$^{-1}$ | 8.841x10$^{-1}$ | 1.697x10$^{-2}$ | 1.972x10$^{-1}$ | α4 |
| $^{289}$Fl | 10.022 | 1.235x10$^{-3}$ | 8.895x10$^{0}$ | 5.379x10$^{0}$ | 1.948x10$^{1}$ | 3.527x10$^{-1}$ | 4.045x10$^{0}$ | SF |

Table 6: Predictions on the mode of decay of $^{306,307}$122 superheavy nuclei and their decay products is given by comparing the alpha half lives and their corresponding spontaneous fission half lives. The calculations are done for zero angular momentum transfers.

| Parent nuclei | $Q_\alpha$ (cal) (MeV) | $T_{SF}$ (s) | $T_{1/2}^{\alpha}$ (s) | | | | | Mode of Decay |
|---|---|---|---|---|---|---|---|---|
| | | | CPPM | CPPMDN | VSS | UNIV | Royer | |
| $^{306}$122 | 12.758 | $2.623 \times 10^{17}$ | $2.071 \times 10^{-4}$ | $1.965 \times 10^{-4}$ | $6.149 \times 10^{-5}$ | $1.932 \times 10^{-5}$ | $3.398 \times 10^{-5}$ | α1 |
| $^{302}$120 | 11.857 | $8.446 \times 10^{9}$ | $7.278 \times 10^{-3}$ | $6.190 \times 10^{-3}$ | $1.810 \times 10^{-3}$ | $4.770 \times 10^{-4}$ | $9.528 \times 10^{-4}$ | α2 |
| $^{298}$118 | 11.175 | $7.221 \times 10^{3}$ | $9.799 \times 10^{-2}$ | $9.448 \times 10^{-2}$ | $2.155 \times 10^{-2}$ | $5.183 \times 10^{-3}$ | $1.096 \times 10^{-2}$ | α3 |
| $^{294}$Lv | 10.224 | $1.336 \times 10^{-1}$ | $1.058 \times 10^{1}$ | $1.052 \times 10^{1}$ | $1.785 \times 10^{0}$ | $3.808 \times 10^{-1}$ | $8.830 \times 10^{-1}$ | SF |
| $^{307}$122 | 13.668 | $6.974 \times 10^{15}$ | $2.074 \times 10^{-6}$ | $2.070 \times 10^{-6}$ | $1.133 \times 10^{-5}$ | $3.563 \times 10^{-7}$ | $3.122 \times 10^{-6}$ | α1 |
| $^{303}$120 | 11.747 | $2.177 \times 10^{8}$ | $1.332 \times 10^{-2}$ | $1.200 \times 10^{-2}$ | $4.199 \times 10^{-2}$ | $8.119 \times 10^{-4}$ | $1.021 \times 10^{-2}$ | α2 |
| $^{299}$118 | 11.095 | $1.804 \times 10^{2}$ | $1.557 \times 10^{-1}$ | $1.224 \times 10^{-1}$ | $4.385 \times 10^{-1}$ | $7.809 \times 10^{-3}$ | $9.754 \times 10^{-2}$ | α3 |
| $^{295}$Lv | 10.404 | $3.234 \times 10^{-3}$ | $2.966 \times 10^{0}$ | $2.141 \times 10^{0}$ | $7.205 \times 10^{0}$ | $1.198 \times 10^{-1}$ | $1.479 \times 10^{0}$ | SF |